\documentclass[english]{aastex}
\usepackage[T1]{fontenc}
\usepackage[latin9]{inputenc}
\usepackage{babel}
\usepackage{float}
\usepackage{textcomp}
\usepackage{amsbsy}
\usepackage{amstext}
\usepackage{amssymb}
\usepackage{graphicx}
\usepackage{wasysym}
\usepackage[unicode=true,pdfusetitle,
 bookmarks=true,bookmarksnumbered=false,bookmarksopen=false,
 breaklinks=true,pdfborder={0 0 0},backref=false,colorlinks=false]
 {hyperref}
\usepackage{breakurl}

\makeatletter

\providecommand{\tabularnewline}{\\}

\newcommand{\areps}{Ann. Rev. Earth Planet. Sci.}
\newcommand{\sci}{Science}
\newcommand{\mps}{Meteor. Planet. Sci.}

\makeatother

\begin{document}

\title{The evolution of asteroids in the jumping-Jupiter migration model}

\author{Fernando Roig}

\affil{Observatório Nacional, Rua Gal. Jose Cristino 77, Rio de Janeiro,
RJ 20921-400, Brazil}

\email{froig@on.br}

\author{David Nesvorn{\'{y}}}

\affil{Southwest Research Institute, 1050 Walnut St., Suite 300, Boulder,
CO 80302, USA}

\email{davidn@boulder.swri.edu}
\begin{abstract}
In this work, we investigate the evolution of a primordial belt of
asteroids, represented by a large number of massless test particles,
under the gravitational effect of migrating Jovian planets in the
framework of the jumping-Jupiter model. We perform several simulations
considering test particles distributed in the Main Belt, as well as
in the Hilda and Trojan groups. The simulations start with Jupiter
and Saturn locked in the mutual 3:2 mean motion resonance plus 3 Neptune-mass
planets in a compact orbital configuration. Mutual planetary interactions
during migration led one of the Neptunes to be ejected in less than
10 Myr of evolution, causing Jupiter to jump by about 0.3 au in semi-major
axis. This introduces a large scale instability in the studied populations
of small bodies. After the migration phase, the simulations are extended
over 4 Gyr, and we compare the final orbital structure of the simulated
test particles to the current Main Belt of asteroids with absolute
magnitude $H<9.7$. The results indicate that, in order to reproduce
the present Main Belt, the primordial belt should have had a distribution
peaked at $\sim10^{\circ}$ in inclination and at $\sim0.1$ in eccentricity.
We discuss the implications of this for the Grand Tack model. The
results also indicate that neither primordial Hildas, nor Trojans,
survive the instability, confirming the idea that such populations
must have been implanted from other sources. In particular, we address
the possibility of implantation of Hildas and Trojans from the Main
Belt population, but find that this contribution should be minor. 
\end{abstract}


\section{Introduction}

The jumping Jupiter model is a model of planetesimal driven migration
of the major planets in which an instability phase arises from the
occurrence of close encounters between the planets. During this phase
of close encounters, a Neptune-size planet is scattered by Jupiter,
and may be ejected from the Solar System. This causes Jupiter's semi-major
axis to ``jump'' inwards by a few tenth of au. The model was first
proposed to solve some limitations of other planetary migration models,
but it has not yet been fully tested against all the possible constraints
imposed by many specific characteristics of the Solar System. The
purpose of this paper is to investigate the effect of the jumping
Jupiter evolution on the dynamics of a primordial belt of asteroids.

Planetesimal driven migration was first proposed by \citet{1984Icar...58..109F}
as the outcome from the gravitational scattering of a remnant disk
of planetesimals by the recently formed major planets. The planets
were initially assumed to be in a more compact orbital configuration.
The exchange of angular momentum between the planets and the planetesimals
induces a smooth change of the planets' semi-major axes resulting
in a divergent migration. Smooth migration models, however, were able
to explain only very few properties of the Solar System (e.g. \citealp{1993Natur.365..819M};
\citealp{1997Sci...275..375L}). Alternatively, instability migration
models have been proposed, where the planetary system evolves through
a short phase of strong dynamical instability. In general, this evolution
does not destabilize the major planets (e.g. Jupiter and Saturn do
not suffer mutual encounters in this model), but it has deep consequences
for the evolution of other Solar System populations. 

The instability model was introduced by \citet{1999Natur.402..635T}
and later reformulated by \citet{2005Natur.435..459T}, becoming known
as the Nice model. In the Nice model, the instability arises from
the crossing of mutual mean motion resonances between Jupiter and
Saturn. The Nice model was able to explain the Late Heavy Bombardment
\citep{2005Natur.435..466G}, and the origin of Trojan asteroids \citep{2005Natur.435..462M},
for example. But it failed to explain other properties, like the orbits
of the terrestrial planets. This led \citet{2009A&A...507.1041M}
and \citet{2009A&A...507.1053B} to propose a revision of the Nice
model, in which the instability arises from the gravitational scattering
of Uranus by Jupiter and Saturn. In this scenario, the successive
close encounters cause sudden changes in Jupiter's (and Saturn's)
semi-major axis, so the model is referred to as the ``jumping Jupiter''
model. A fundamental shortback of the original jumping Jupiter evolution
is that, more often than not, Uranus or Neptune is ejected from the
Solar System. More recently, \citet{2011ApJ...742L..22N} and \citet{2012AJ....144..117N}
attempted to solve this limitation considering systems with five initial
planets: Jupiter, Saturn, and three ice giants. The fifth hypothetical
planet is then ejected, leaving a system as we known today. This model
has had a great success in explaining the origin of the irregular
satellites of Jupiter \citep{2014ApJ...784...22N}, several properties
of the Jovian satellites \citep{2014AJ....148...25D,2014AJ....148...52N},
the origin of Jupiter Trojans \citep{2013ApJ...768...45N}, and the
structure of the Kuiper belt \citet{2015arXiv150606019N}. It has
also proven to fulfill the constraints from the terrestrial planets
and the architecture of the outer planets. But it has not yet been
tested for the asteroid belt.

In contrast with many other populations of minor bodies, the main
advantage of the asteroid belt is that it constitutes a very well
known population. Its orbital distribution is little affected by observational
biases, its size distribution is complete up to very small sizes,
and its taxonomical distribution is well understood. This provides
a number of tight constraints on the early evolution of the Solar
System. In particular, we know that the asteroid belt should have
suffered a significant primordial excitation, a significant mass depletion,
and a significant mixture of taxonomies. Attempts to explain these
issues were first due to \citet{1992Icar..100..307W}, \citet{2001Icar..153..338P},
and \citet{2007Icar..191..434O}. For a detailed review, we refer
the reader to \citet{2015arXiv150106204M}. Two results are of particular
relevance to the present work.

First, Walsh et al. \citeyearpar{2011Natur.475..206W,2012M&PS...47.1941W}
proposed that the mass depletion and the primordial mixing of taxonomies
in the asteroid belt could be the consequence of the early migration
of Jupiter and Saturn in a gas disk \citep{2001MNRAS.320L..55M,2014ApJ...795L..11P}.
This evolution, which presumably happened prior to the phase of planetesimal
driven migration, became known as the Grand Tack. The time elapsed
between the end of the Grand Tack (i.e. end of the gas driven migration)
and the occurrence of the dynamical instability is poorly constrained.
It could be as short as a few Myr and as long as $\sim500$ Myr, the
latter assuming that the instability was responsible for triggering
the Late Heavy Bombardment about 4 Gyr ago. Since our results have
implications for the Grand Tack model, we will come back to this issue
later.

Second, Minton and Malhotra \citeyearpar{2009Natur.457.1109M,2011ApJ...732...53M}
addressed the stirring of the asteroids eccentricities and inclinations
during the phase planetesimal driven migration. They found that in
a smooth migration scenario, the characteristic time of migration
should be unrealistically fast in order to let resonances sweeping
to account for the orbital excitation of the asteroids. This result
favors instability models \emph{alla} jumping Jupiter, that could
effectively provide the fast migration rates required. In particular,
\citet{2010AJ....140.1391M} applied a jumping Jupiter model with
only four planets to the asteroid main belt, and found that it is
possible to reproduce its current distribution in inclinations provided
that the original distribution had an upper cutoff around $20^{\circ}$.
We will discuss this result later in the light of our findings.

Our goal here is to analyze and discuss the significance of the jumping
Jupiter migration in the evolution of the asteroid belt. We perform
simulations that reproduce the current orbital distribution of the
belt and allow us to constrain its initial distribution, that is,
at the beginning of the planetesimal driven migration. The paper is
divided as follows: in Sect. \ref{sec:Simulations}, we describe the
methodology used in our simulations. In Sect. \ref{sec:Results} we
present our results, discussing the implications for the asteroid
main belt, the Grand Tack model, and the Hilda and Trojan populations.
Finally, in Sect. \ref{sec:Conclusions} we summarize our conclusions.

\section{Simulations\label{sec:Simulations}}

We have run a series of numerical simulations of the evolution of
the asteroid belt from the epoch before the triggering of the planetesimal
driven migration to the present days. The asteroid belt is represented
by a large number of test particles perturbed by the Jovian planets.
It is worth stressing that the terrestrial planets have not been included
in our simulations. The initial orbital distribution of the test particles
is uniform, and the resulting final distribution is compared to the
presently observed asteroid belt. This allows us to remap the initial
distribution onto any desired non uniform distribution and to determine
the one that provides the best fit to the current asteroid belt. We
have assumed a strategy consisting in dividing the whole evolutionary
process into four stages or phases, that we describe in the following.

\subsection{Phase 0: Before the instability\label{sub:Phase-0}}

This phase is intended to simulate the evolution of a primordial asteroid
belt between the end of the gas driven migration (Grand Tack) and
the occurrence of the jumping Jupiter instability. During this phase,
the planetary system is constituted by Jupiter, Saturn and three Neptune-size
ice giants. The planets do not migrate and stay in a mutual resonant
configuration that is the outcome of the previous gas driven migration
phase \citep{2014ApJ...795L..11P}. In particular, Jupiter and Saturn
are locked in the 3:2 mean motion resonance, with Jupiter slightly
outside of its present orbit. The initial osculating values of semi-major
axis $a$, eccentricity $e$, and inclination $I$ are shown in Table
\ref{tab:Planetary-initial-conditions}. This configuration is stable
over the simulation time span of 500 Myr. 

\begin{figure}[H]
\protect\caption{Distribution of the test particles at the beginning of Phase 0. Black
dots represent the main belt (MB) population. Red dots represent the
Hilda group (HG) centered at $\sim4.1$ au, and the Trojan swarms
(TS) centered at $\sim5.4$ au. Note that the positions of these populations
are moved outwards with respect to their present locations, since
the 3:2 and 1:1 mean motion resonances with Jupiter are shifted to
larger semi-major axes. The big black dot represents the initial position
of Jupiter. }
\label{init-fase0}

\centering{}\medskip{}
\includegraphics[width=1\textwidth]{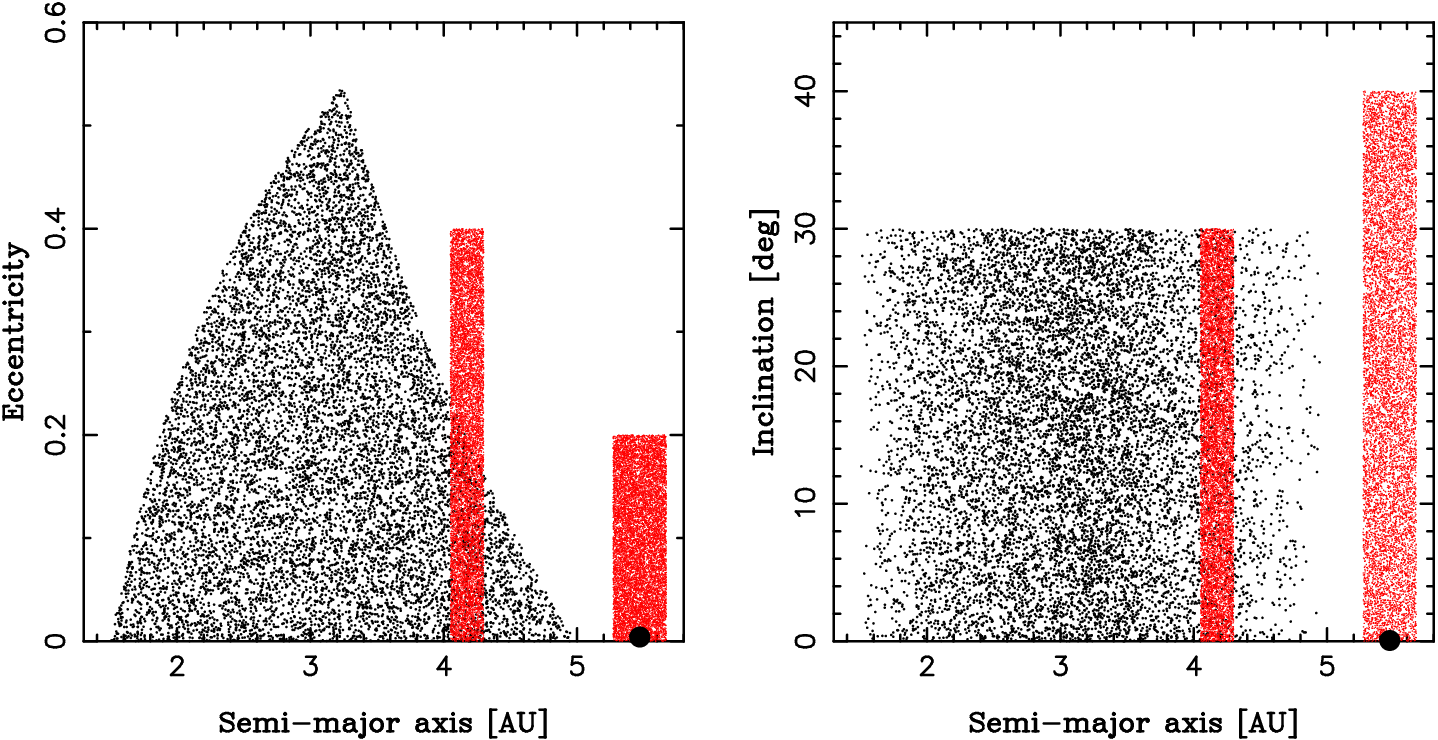}
\end{figure}

The test particles are grouped into three populations: the Main Belt,
the Hilda group and the Trojan swarms. Each population is represented
by 10,000 initial conditions uniformly (randomly) distributed over
the following intervals:
\begin{itemize}
\item $1.5\leq a\leq5.0$ au, perihelion distance $q>1.5$ au, aphelion
distance $Q<5.0$ au, $I<30^{\circ}$, for the Main Belt (hereafter
MB set);
\item $4.0\leq a\leq4.3$ au, $e<0.4$, $I<30^{\circ}$, for the Hilda group
(hereafter HG set);
\item $5.2\leq a\leq5.7$ au, $e<0.2$, $I<40^{\circ}$, for the Trojan
swarms (hereafter TS set).
\end{itemize}
The initial angles $M,\omega,\Omega$ (mean anomaly, argument of perihelion,
and longitude of node) are randomly distributed between 0 and $360^{\circ}$
in all cases. Figure \ref{init-fase0} shows the initial distribution
of the test particles. The simulations are carried out using the SWIFT\_RMVS4
symplectic integrator\footnote{\url{https://www.boulder.swri.edu/~hal/swift.html}},
with a time step of 0.05 yr. The test particles are eliminated if
they approach within the radius of any planet, or if $q$ gets smaller
than 1 au, or if $a$ evolves beyond 100 au. 

Table \ref{statistic-of-surviving} shows the number of surviving
test particles at two specific times: 10 Myr and 500 Myr. We have
chosen these times because they represent the state of the asteroid
belt in two interesting limits: (i) when the instability is triggered
shortly after the end of the gas driven migration, and (ii) when the
instability happens at a time compatible with the Late Heavy Bombardment.
We note that at least half of the MB population is stable, while the
HG and TS populations are significantly eroded\footnote{It is worth noting that, due to the partial overlap of the MB and
HG sets, the actual number of survivors in the Hilda region is about
1.5 times larger than in the HG set alone. }. This reinforces the idea that these latter groups are not primordial
but implanted from other sources (e.g. \citealp{2009Natur.460..364L};
\citealp{2013ApJ...768...45N}). An analysis of the reason for this
erosion is beyond the scope of this paper (see \citealp{2002mcma.book.....M};
\citealt{2009MNRAS.399...69R}).

\subsection{Phase 1: Jumping Jupiter\label{sub:Phase-1}}

\begin{figure}[H]
\protect\caption{The three cases of jumping Jupiter instability from \citet{2013ApJ...768...45N}
that are studied in this paper. For each planet, the plots show the
semi-major axis and the perihelion and aphelion distances. Black corresponds
to Jupiter, red to Saturn, and green, blue and cyan to the three icy
giants.}
\label{cases-mig}

\centering{}\medskip{}
\includegraphics[width=0.33\textwidth]{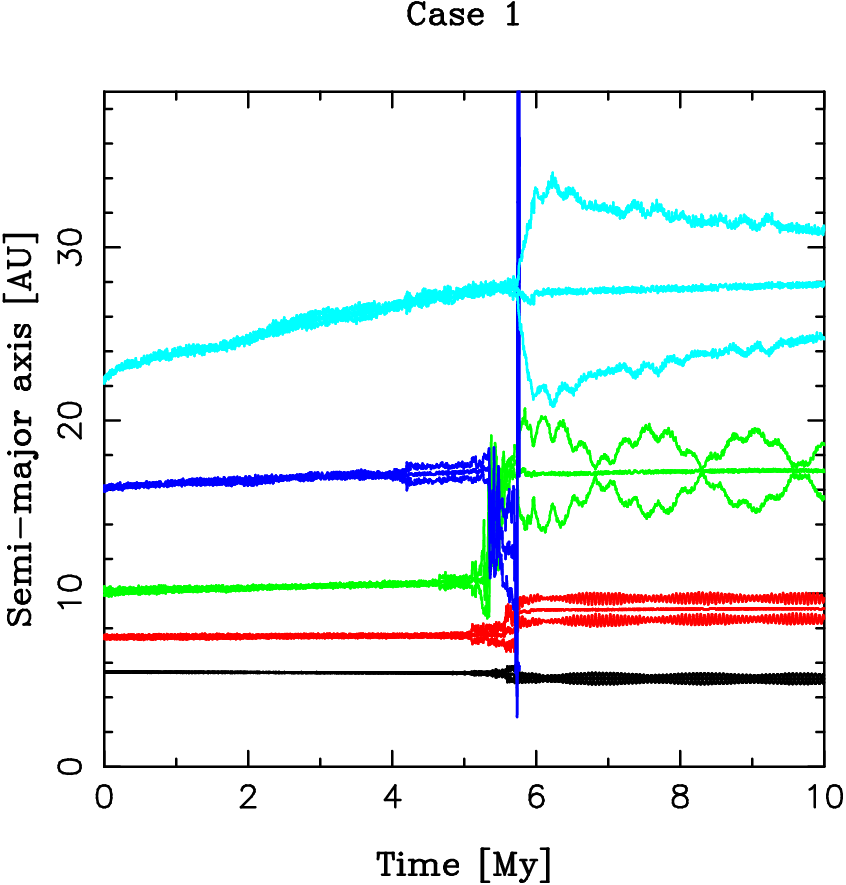}%
\includegraphics[width=0.33\textwidth]{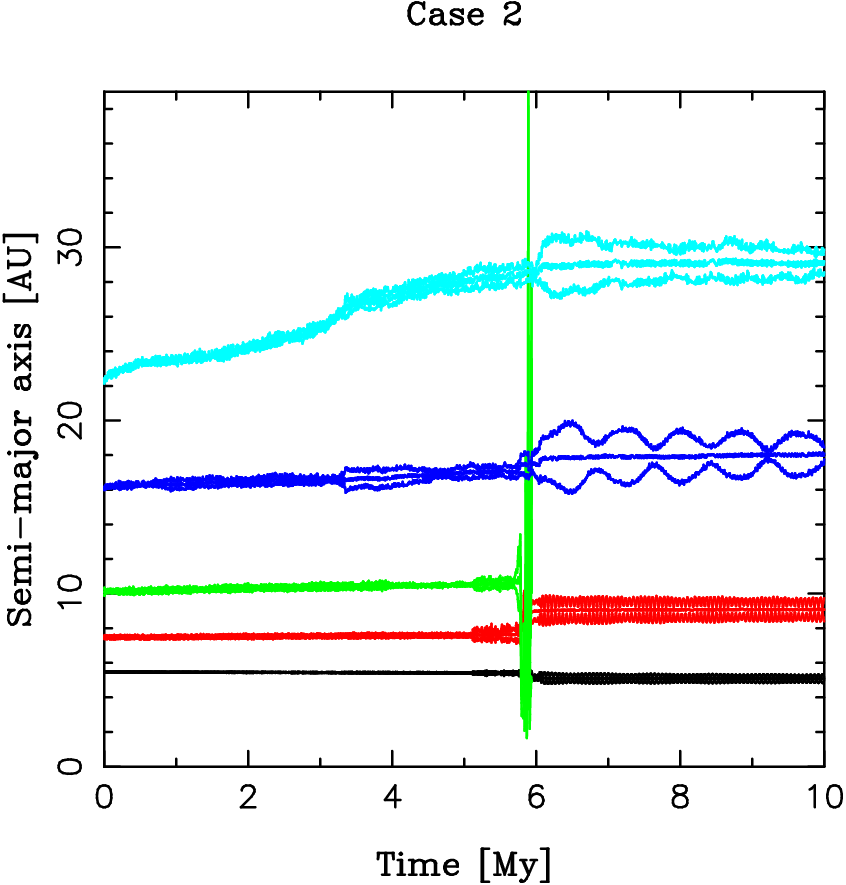}%
\includegraphics[width=0.33\textwidth]{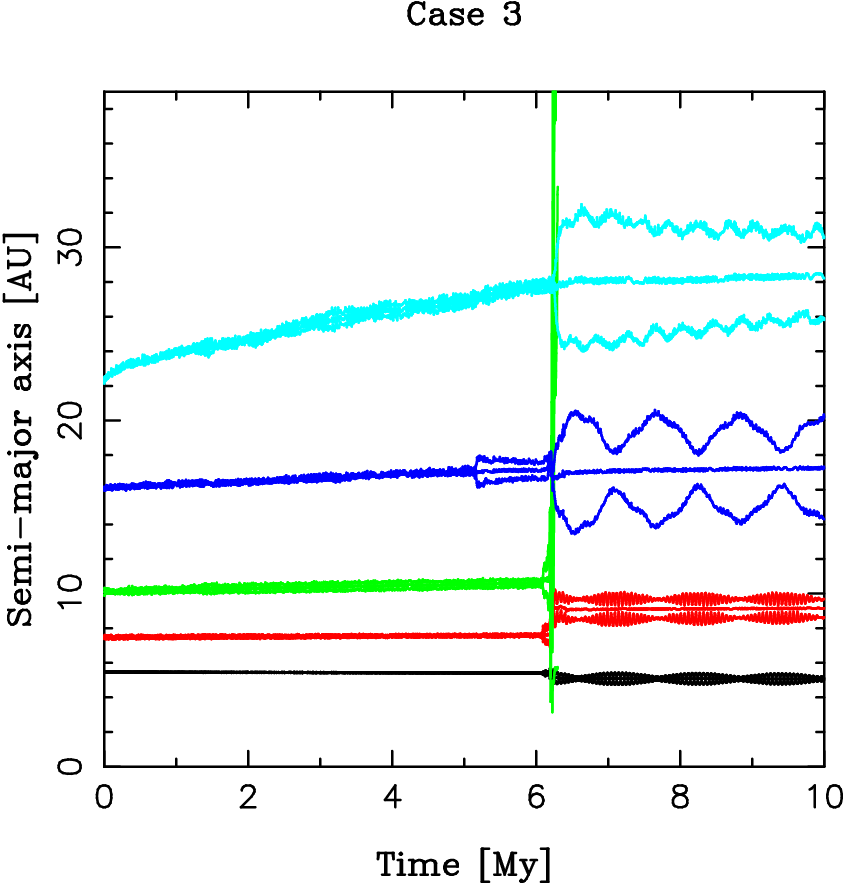}
\end{figure}

The purpose of this phase is to simulate the epoch of the jumping
Jupiter instability. During this phase, the five Jovian planets migrate
according to three specific evolutions previously developed by \citet{2012AJ....144..117N}.
These authors performed realistic simulations of migration, where
the planets interact with a massive disk of planetesimals initially
located beyond the outermost planet. In the simulations, the planets'
positions and velocities were stored in a file at 1 yr intervals over
a total time span of 10 Myr. We have not reproduced the simulations
of \citet{2012AJ....144..117N}; rather we have mimicked the migration
by reading the stored positions and interpolating them using an approach
described in \citet{2013ApJ...768...45N}. Figure \ref{cases-mig}
shows the three evolutions we have considered, which we refer to as
Cases 1, 2 and 3, respectively. The main differences between the three
cases are:
\begin{itemize}
\item The Ice \#2 planet (middle ice giant) is ejected in Case 1, while
the Ice \#1 (innermost ice giant) is ejected in Cases 2 and 3;
\item Case 2 produces a large number of encounters between Jupiter and the
ejected planet, which makes Jupiter to suffer a lot of short jumps.
Cases 1 and 3, on the other hand, produce less interactions and make
Jupiter to experience a few longer jumps;
\item In Case 1, the ejected planet reaches minimum heliocentric distances
of $\sim2$ au during the instability, which means that it penetrates
deeply in the asteroid belt. In Case 2, the heliocentric distance
of the ejected planet gets to $\sim1.5$ au very briefly. In Case
3, the heliocentric distances do not get much below $\sim3$ au.
\end{itemize}
We recall that in all three cases, the planets are initially in the
same orbital configuration given in Table \ref{tab:Planetary-initial-conditions}.
The different evolutions in each case arise from the different parameters
of the planetesimal disk considered by \citet{2012AJ....144..117N}.
The net inwards jump of Jupiter is $\sim0.3$ au in all cases, and
the instability always occurs between 5.5 and 6.5 Myr after the start
of the simulations.

These three instability cases have been previously tested against
a number of constraints. They satisfy the constraints defined in \citet{2012AJ....144..117N},
namely, the final orbits of the outer planets are similar to the real
orbits. For example, the proper mode in Jupiter's orbit is excited
to its present value by planetary encounters. All three cases also
satisfy the terrestrial planets constraint in that Jupiter's orbit
discontinuously evolves during planetary encounters \citep{2009A&A...507.1053B}.
This is needed to avoid secular resonances with the terrestrial planets,
which would otherwise lead to a disruption of the terrestrial planets
system. All three cases are equally good in explaining the capture
and orbital distribution of Jupiter Trojans, including their high
orbital inclinations \citep{2013ApJ...768...45N}. Case 2 shows a
richer history of planetary encounters, which leads to a large perturbation
of the Galilean satellite orbits that is difficult to reconcile with
the present orbits of these moons \citep{2014AJ....148...25D}. Cases
1 and 3, on the other hand, have fewer planetary encounters and satisfy
the Galilean satellites constraint. 

The initial conditions of the test particles for Phase 1 have been
cloned from the same initial conditions used for Phase 0. The reason
for this is twofold: (i) we want to preserve the initial angular phases
between the planets and the test particles, which is especially critical
for the HG and TS populations, and (ii) we also want to increase the
number statistics. Note, however, that we have only cloned the initial
conditions of those test particles that survived Phase 0 after 10
Myr and 500 Myr, respectively. This produces two different sets of
initial conditions for Phase 1. We refer to them as the ``Early Instability''
(EI) set and the ``Late Instability'' (LI) set, respectively. 

\begin{figure}[H]
\protect\caption{Initial conditions for Phase 1: Early Instability set (left) and Late
Instability set (right). Black dots correspond to the MB population.
Red dots correspond to the HG and TS populations (only in the Early
Instability set). The big black dot represents Jupiter}
\label{init-fase1}

\centering{}\medskip{}
\includegraphics[width=1\textwidth]{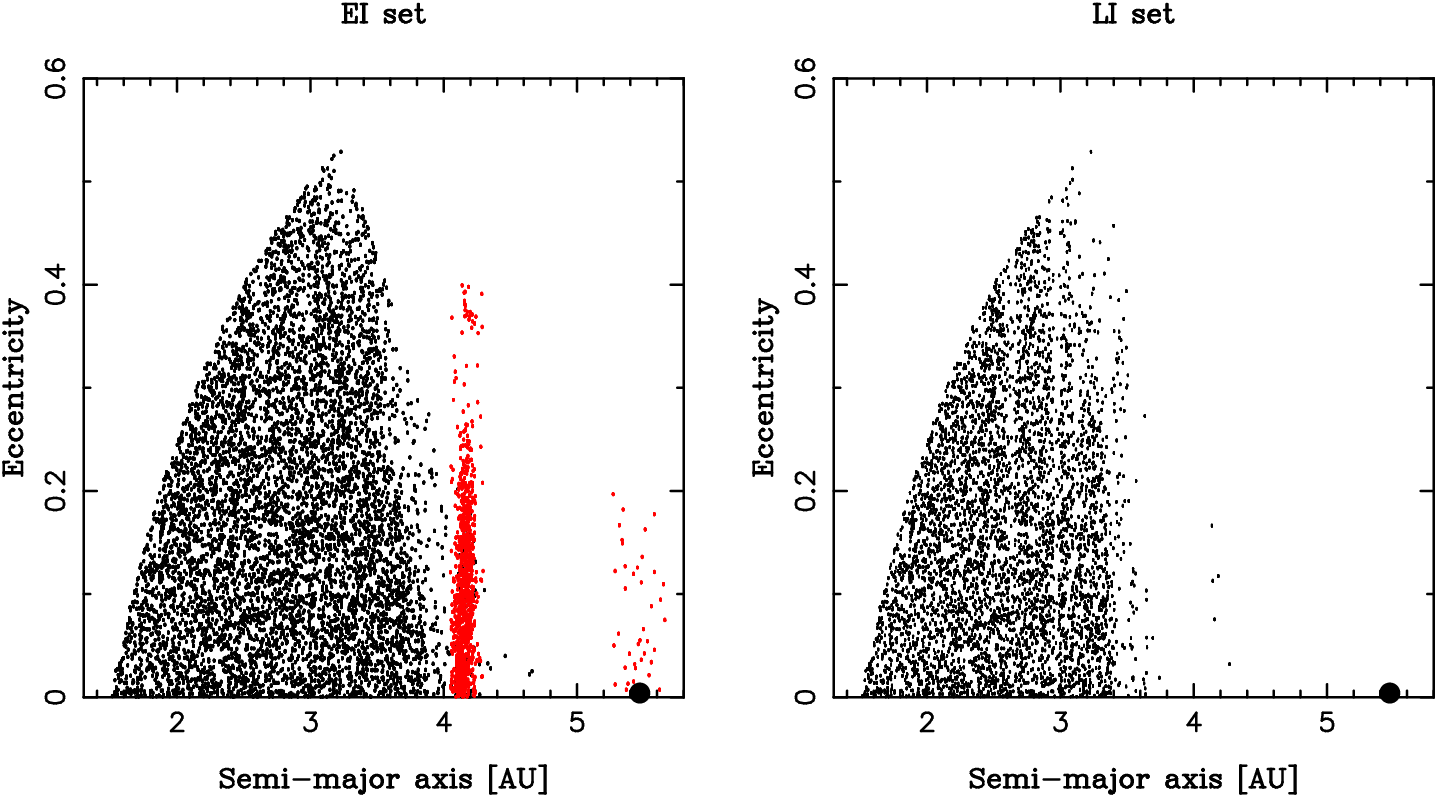}
\centering{}\medskip{}
\includegraphics[width=1\textwidth]{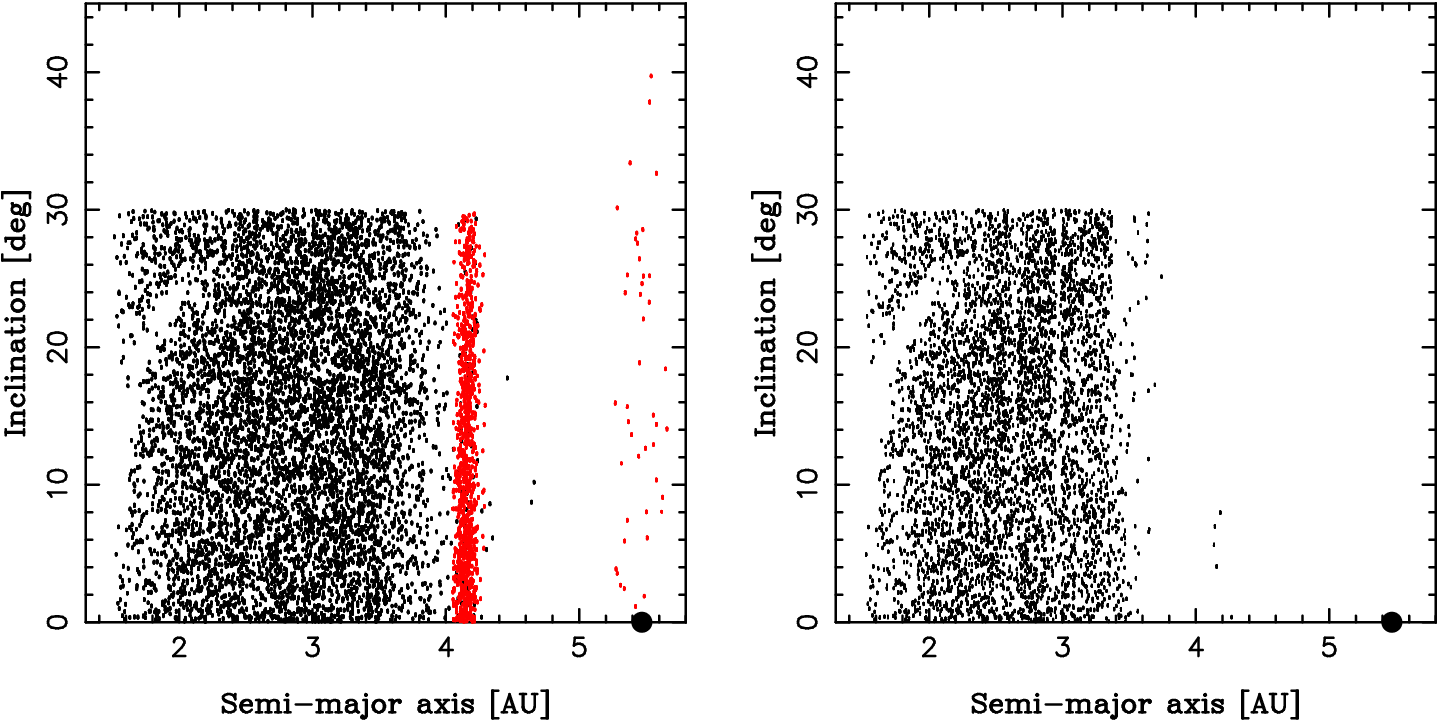}
\end{figure}

For the EI set, each initial condition of Phase 0 that survived 10
Myr has been randomly cloned 15 times within an interval $\Delta a=\pm0.001$
au, $\Delta e=\pm0.001$ and $\Delta I=\pm0.1^{\circ}$ around the
reference orbit. The remaining orbital elements were set equal to
those of the reference orbit. After the cloning process, we have ended
up with a MB population of $\sim102~000$ test particles, and a HG+TS
population of $\sim15~000$ test particles. For the LI set, we have
applied a similar method, but only to the MB population. We have not
taken into account the HG and TS populations due to the small amount
of survivors at 500 Myr. We have created 21 clones of each initial
condition of Phase 0 that survived after 500 Myr, ending up with a
MB population of $\sim99~000$ test particles. Figure \ref{init-fase1}
shows the initial distribution of the EI and LI sets. The main difference
between the two sets is the larger depletion at $a>3.4$ au observed
in the LI set.

The simulations have been carried out using a modified version of
the SWIFT\_RMVS3 symplectic integrator, that interpolates the stored
planetary positions and propagates the test particles (cf. \citealp{2013ApJ...768...45N}).
The total time span of this phase is 10 Myr, and the integration time
step is 0.05 yr. The test particles are discarded if they hit any
planet, or if $q<1$ au or $a>100$ au.

\subsection{Phase 2: Residual migration}

\begin{figure}[H]
\protect\caption{The distribution of the MB population in the Early Instability set,
at the end of Phase 1 (migration Case 1). About $\sim46\,000$ test
particles (of the originally 102~000) survived the instability. Their
orbits are used as the initial conditions for Phase 2. The big dot
represents Jupiter. The gap at $\sim2$ au is due to the $\nu_{6}$
secular resonance.}
\label{init-fase2}

\centering{}\medskip{}
\includegraphics[width=1\textwidth]{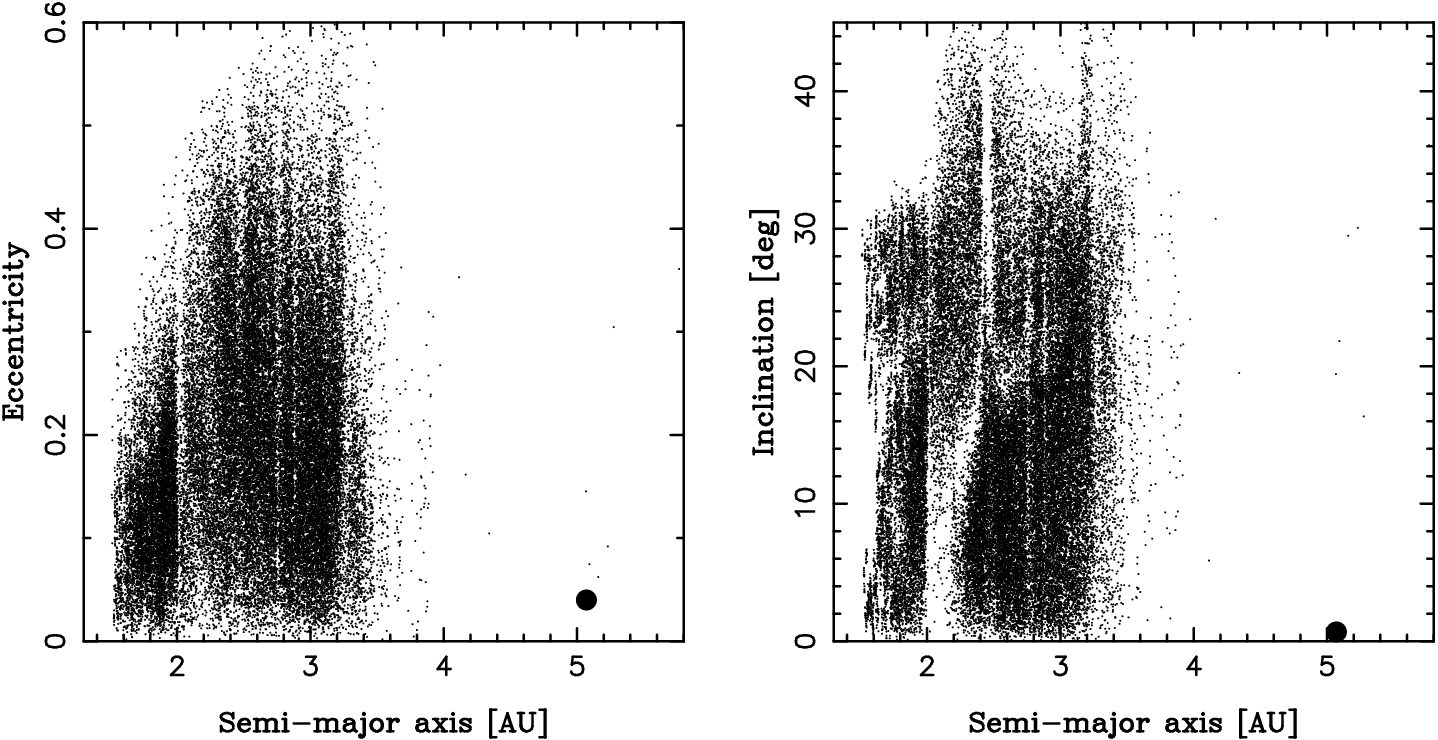}
\end{figure}

This phase is intended to simulate the residual migration of the Jovian
planets after the jumping Jupiter instability. The residual migration
is a smooth process caused by the interaction of the planets with
the disk of planetesimals. It drives the planets to reach their current
orbits, making Jupiter to migrate inwards, while Saturn and the two
remnant ice giants (Uranus and Neptune) migrate outwards. This migration
has been simulated using a modified version of the SWIFT\_RMVS4 code,
that applies a non conservative acceleration to each planet. The non
conservative acceleration $\boldsymbol{a}$ has the form:
\begin{equation}
\boldsymbol{a}=\alpha\exp\left(-\frac{t}{\tau}\right)\boldsymbol{v}-2\eta\frac{\boldsymbol{r}\cdot\boldsymbol{v}}{r^{2}}\boldsymbol{r}\label{eq:accel}
\end{equation}
where $t$ is the time, $\boldsymbol{r},\boldsymbol{v}$ are the position
and velocity vector of the planet, and $\alpha,\tau,\eta$ are constants
specific to each planet. The first term provides a smooth drift in
semi-major axis, while the second term produces damping in eccentricity.
An additional term of the form $-2\zeta v_{z}$ has been introduced
in the $z$ component of the acceleration to produce inclination damping.
The values of $\alpha,\tau,\eta,\zeta$ were tuned to get each planet
in its present orbit at the end of the simulation. The initial conditions
for both the planets and the test particles have been taken directly
from the final conditions produced by Phase 1. An example of the initial
conditions is shown in Fig. \ref{init-fase2}. The most notable feature
in this figure is the gap around 2 au that has been opened by the
$\nu_{6}$ secular resonance; this resonance is located more or less
at its present location \citep{1991Icar...93..316K} already at the
end of Phase 1.

As in previous phases, the test particles are discarded if they hit
any planet, or if $q<1$ au or $a>100$ au. The total time span of
this phase is 100 Myr, and the integration time step is 0.05 yr.

\subsection{Phase 3: Long term evolution}

This is the last stage of our simulations. It intends to reproduce
the evolution of the asteroid belt from the end of the planetesimal
driven migration to the present days. The planets do not migrate any more,
and their initial conditions are those resulting from Phase 2. The
total time span of this phase is 4 Gyr. The evolution has been simulated
using the SWIFT\_RMVS4 code, with a time step of 0.05 yr. 

The initial conditions for the test particles have been taken from
the final conditions produced by Phase 2. However, at the end of Phase
2 there are typically between 25~000 and 50~000 surviving test particles,
depending on the migration case and on the EI or LI set. Since it
is not possible to simulate such amount of test particles over 4 Gyr
in a reasonable CPU time, we have performed a down-sampling of the
sets. The down-sampling process involves the following steps:
\begin{itemize}
\item test particles with $q<1.6$ au and $a<2.1$ au are not considered
at all. This is because we are not interested at this time in the
Mars crossing population ($q<1.6$ au), nor in the particles that
survive in the Extended Inner Belt ($a<2.1$ au; \citealp{2012Natur.485...78B}).
This procedure typically removes $\sim30$\% of the test particles
from the whole sample. In principle, these test particles should have
been significantly depleted during previous phases if the terrestrial
planets were included in the simulations. Actually, the only remnant
of this population that we recognize today is the Hungaria group,
whose analysis is beyond the scope of the present work\footnote{The Hungaria group is above the absolute magnitude limit imposed to
our comparison population in Sect. \ref{sec:Results}.}.
\item test particles with $3.5\leq a\leq5.3$ au are all considered. This
is because this region is significantly depleted during the previous
phases and we want to keep track of the few survivors, especially
at the HG and TS population (see Sect. \ref{sub:The-Hilda-and}).
Typically, these test particles represent less than 0.3\% of the whole
sample. 
\item the remaining test particles with $a<3.5$ au are selected at random,
producing a down-sampled set with only 1~008 initial orbits. This
implies that the set is down-sampled by a factor $f_{\mathrm{down}}$
between 15 and 35, depending on the migration case and on the EI or
LI set.
\end{itemize}
Note that this approach assigns a larger weight to test particles
with $a\geq3.5$ au than to the rest of the simulated particles. Nevertheless,
this does not introduce any appreciable bias in our results, because
these particles contribute very little to the distribution of the
down-sampled set (see Sect. \ref{sec:Results}). 

\begin{figure}[H]
\protect\caption{The tiny dots represent the distribution of the MB population in the
Early Instability set, at the end of Phase 2 (migration Case 1). About
$\sim32\,000$ test particles (of the initially 46~000) survived
the residual migration. Test particles with $q<1.6$ au and $a<2.1$
au (tiny black dots) are not taken into account. Of the remaining
$\sim22\,000$ test particles (tiny cyan dots), we perform a random
down-sampling to get the set of $1\,008$ particles represented by
the blue dots (note that particles with $a\geq3.5$ au have not been
down-sampled). These are used as initial conditions for Phase 3. In
red, we indicate the location of the main mean motion resonances with
Jupiter (left), and the location of the main secular resonances (right).}
\label{init-fase3}

\centering{}\medskip{}
\includegraphics[width=1\textwidth]{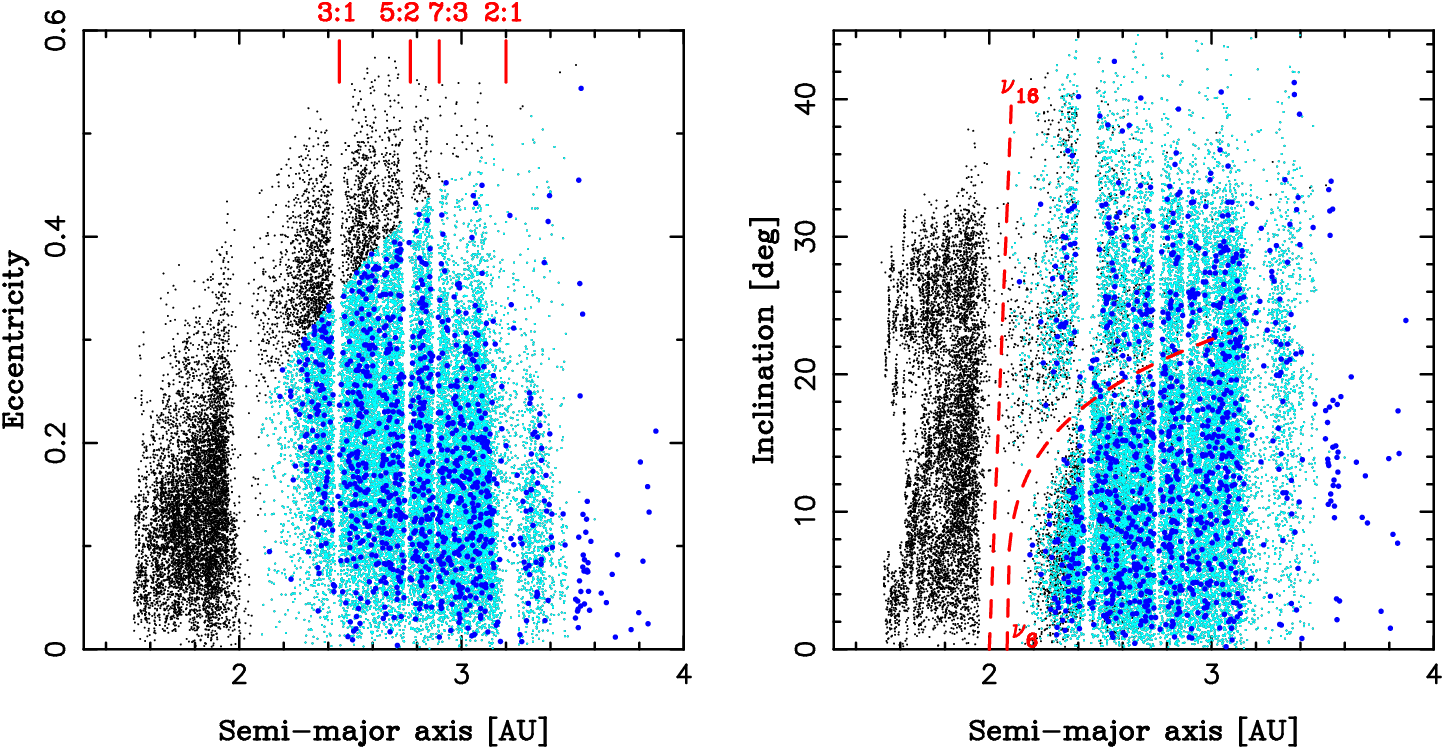}
\end{figure}

An example of the down-sampling procedure is shown in Fig. \ref{init-fase3}.
Looking at the cyan dots in this figure, we can see that the main
features of the present main belt have been already sculpted at the
end of Phase 2. In particular, we can clearly see the gaps opened
by the $\nu_{6}$ and $\nu_{16}$ secular resonances (indicated by
dashed lines), as well as the Kirkwood gaps opened at the 3:1, 5:2.
7:3 and 2:1 mean motion resonances with Jupiter (indicated at the
top of the plot). 

During Phase 3, each initial set of 1~008 test particles has been
propagated together with the four Jovian planets. Depending on the
migration case and the EI or LI set, between 15\% and 25\% of the
particles become discarded by the same reasons as in the previous
phases.

\section{Results\label{sec:Results}}

We have compared the final state of the test particles at the end
of Phase 3 with the current distribution of the real asteroid belt.
We have taken into account the 574 asteroids with $q\geq1.6$ au,
$a\leq4.2$ au, and absolute magnitude $H\leq9.7$, that corresponds
to diameters larger than 40-70 km depending on the assumed albedo.
The magnitude limit is the same considered by \citet{2010AJ....140.1391M}.
The upper cutoff in $a$ has been adopted to exclude the Trojan population
from the comparison. These asteroids may be considered ``primordial''
in the sense that they have not suffered catastrophic collisions over
the age of the Solar System \citep{1992Icar...97..111F}. They have
not suffered significant variations of their orbital elements either,
especially in terms of semi-major axis, because they are too big to
be affected by the Yarkovsky effect \citep{2006AREPS..34..157B}.
Moreover, these asteroids constitute an observationally complete sample.

\begin{figure}[H]
\protect\caption{The blue dots represent the final distribution of 800 test particles
in the MB population (Early Instability set, migration Case 1), at
the end of Phase 3. The distribution of 574 real asteroids with $H\leq9.7$
(red dots) is shown for comparison. The dotted-dashed line in panel
(a) corresponds to $q=1.6$ au. The dashed lines in panel (b) give
the approximate location of the $\nu_{6}$ and $\nu_{16}$ secular
resonances.}
\label{comparison}

\centering{}\medskip{}
\includegraphics[width=1\textwidth]{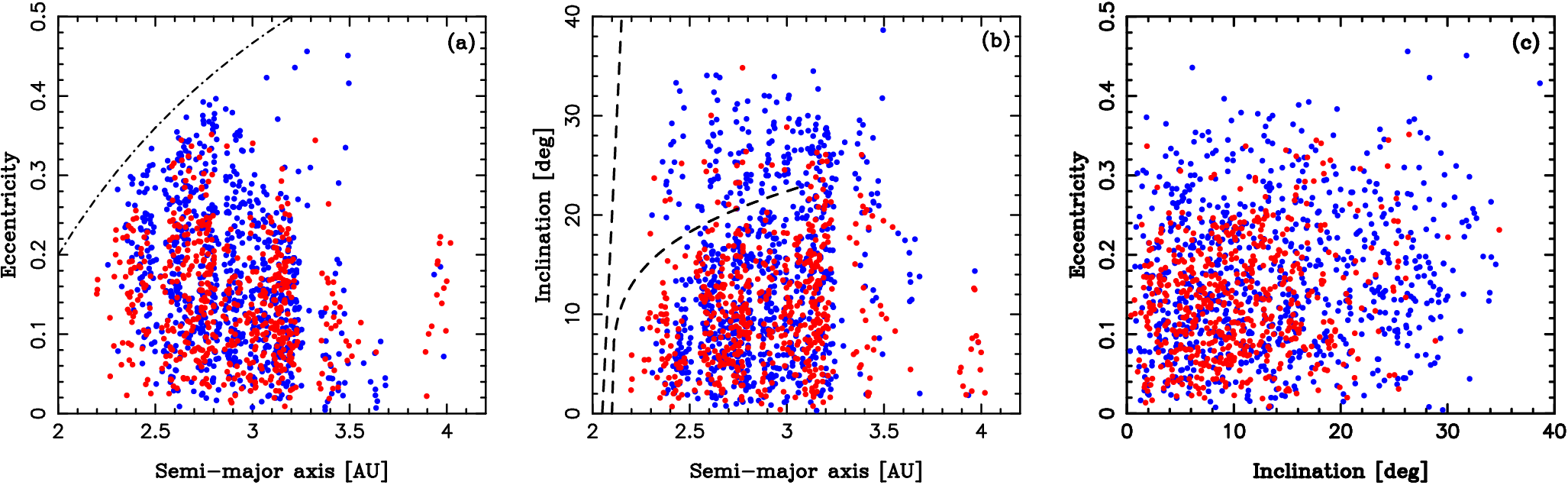}
\end{figure}

\begin{figure}[H]
\protect\caption{Cumulative distributions in $e$ and $I$, and histogram of the density
distribution in $a$, comparing the real asteroids with $H\leq9.7$
(full lines) to the simulated test particles at the end of Phase 3
(dashed lines and gray histogram). The simulation is the same presented
in Fig. \ref{comparison}.}
\label{distribs}

\centering{}\medskip{}
\includegraphics[width=1\textwidth]{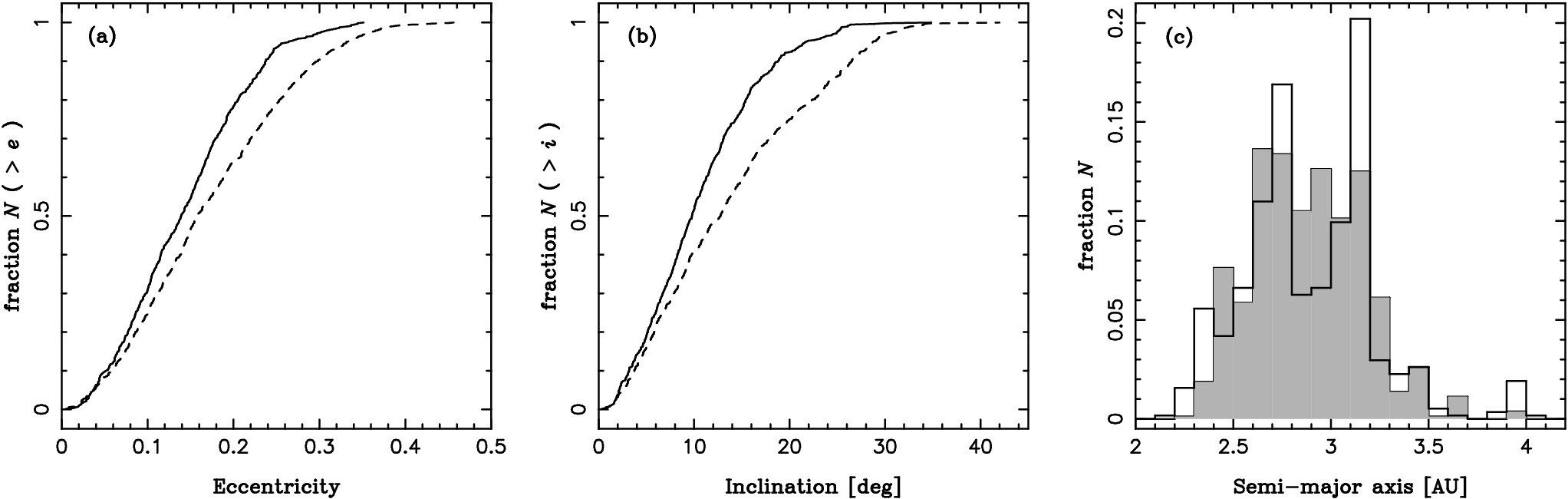}
\end{figure}

Figure \ref{comparison} shows the distribution of the 574 real asteroids
together with the distribution of the 800 test particles with $q\geq1.6$
au and $a\leq4.2$ au from the EI set, that survived at the end of
Phase 3 in migration Case 1\footnote{It is worth noting that, at the end of Phase 3, Jupiter is at $a\sim5.1$
au, that is a little bit inwards than the actual location ($a\simeq5.2$
au). This produces a shift of all the resonance locations (Kirkwood
gaps) in our simulated test particles. To correct this, we have applied
an appropriate shift in the semi-major axis of the particles to get
the Kirkwood gaps in their right position. }. We can see a quite good general agreement between the two distributions.
A more detailed way to compare the results is to look at the cumulative
distributions in $e$ and $I$, and the density distribution in $a$.
This is shown in Fig. \ref{distribs}. As stated before, we have verified
that the region between 3.5 and 4.2 au does not contribute significantly
to the overall statistics. In fact, this region accounts for only
3\% of the real population considered here, and between 0.5\% and
3\% of the test particles (depending on the migration case and the
EI or LI sets).

In Fig. \ref{distribs}, we note that there is a clear excess of high
eccentricity ($e>0.25$) and high inclination ($I>20^{\circ}$) test
particles with respect to the real asteroid belt. In particular, we
have verified that the high inclination particles had high inclinations
already at the beginning of Phase 0. Therefore, this excess could
be eliminated, for example, by applying a suitable upper cutoff to
the initial population. More realistic initial distributions can be
tested to find the one that best matches the real asteroid belt, as
we show in the following.

Our first goal has been to match the cumulative distributions in $e$
and $I$; the differential distribution in $a$ is considered later
(Sect. \ref{sub:The-distribution-in}). The procedure involved to
remap the uniform initial orbital distribution of the test particles
into a non uniform distribution by selecting appropriate initial orbits.
Then, the final state of these selected orbits was used to match the
observations. For this purpose, we have tested four possible types
of non uniform initial density distributions:
\begin{itemize}
\item upper cutoff distribution
\begin{equation}
p(x)=\left\{ \begin{array}{c}
1\qquad\mathrm{if}\,x\leq c\\
0\qquad\mathrm{if}\,x>c
\end{array}\right.\label{eq:cut}
\end{equation}
parametrized by the cutoff $c$ (with $x\equiv e,I$ );
\item Rayleigh distribution
\begin{equation}
p(x)=\frac{x}{\gamma^{2}}\exp\left(-\frac{x^{2}}{2\gamma^{2}}\right)\label{eq:ray}
\end{equation}
parametrized by the mode $\gamma$;
\item Maxwell distribution
\begin{equation}
p(x)=\sqrt{\frac{2}{\pi}}\frac{x^{2}}{\beta^{2}}\exp\left(-\frac{x^{2}}{2\beta^{2}}\right)\label{eq:max}
\end{equation}
parametrized by the mode $\sqrt{2}\beta$; and 
\item Gaussian distribution
\begin{equation}
p(x)=\frac{1}{\sigma\sqrt{2\pi}}\exp\left(-\frac{(x-\mu)^{2}}{2\sigma^{2}}\right)\label{eq:gaus}
\end{equation}
parametrized by the mean $\mu$ and the standard deviation $\sigma$.
\end{itemize}
Assuming a given type of initial distribution for $e$, and a given
type of initial distribution for $I$ (not necessarily the same),
we have scanned all the possible values of the distribution parameters.
Then, we have applied a Kolmogorov-Smirnoff (K-S) test to compare
the final cumulative distributions in $e$ and $I$ obtained for each
value of the parameters, with the real asteroid belt. The results
are presented in the form of color maps, where the color level indicates
the value of the K-S statistic $D_{\mathrm{KS}}$. In these maps,
the horizontal axis gives the value of the parameter for the assumed
initial distribution in $e$. The vertical axis gives the value of
the parameter for the assumed initial distribution in $I$. The white
curve encloses the region where $D_{\mathrm{KS}}$ has the highest
significance level \citep{1992NumRecip}. We recall that in the case
of upper cutoff, Rayleigh and Maxwell distributions, there is only
one parameter to vary. In the case of the Gaussian distribution, we
have fixed one of the parameters (either $\mu$ or $\sigma$) and
let the other vary.

\subsection{The Early Instability set}

\begin{figure}[H]
\protect\caption{Comparison of the 2-D distribution in the $e,I$ plane between the
test particles of the EI set at the end of Phase 3 and the current
asteroid belt. The color scale gives the values of the two dimensional
Kolmogorov-Smirnoff statistic $D_{\mathrm{KS}}$. Best fits are obtained
for $D_{\mathrm{KS}}\leq0.1$ (dark blue color). The white curve encloses
the region where $D_{\mathrm{KS}}$ has the highest significance levels.
The black pixels in the lower left corner of the plots correspond
to values for which the K-S test could not be applied due to the small
amount of data ($N<5$). The abscissas give the parameter value of
the assumed initial distribution in $I$ (i.e. at the beginning of
Phase 1). The ordinates give parameter value of the assumed initial
distribution in $e$. The top row corresponds to initial upper cutoff
distributions. The bottom row corresponds to initial Rayleigh distributions.
Each column corresponds to a different migration case.}
\label{ks-test-early}

\centering{}\medskip{}
\includegraphics[width=0.33\textwidth]{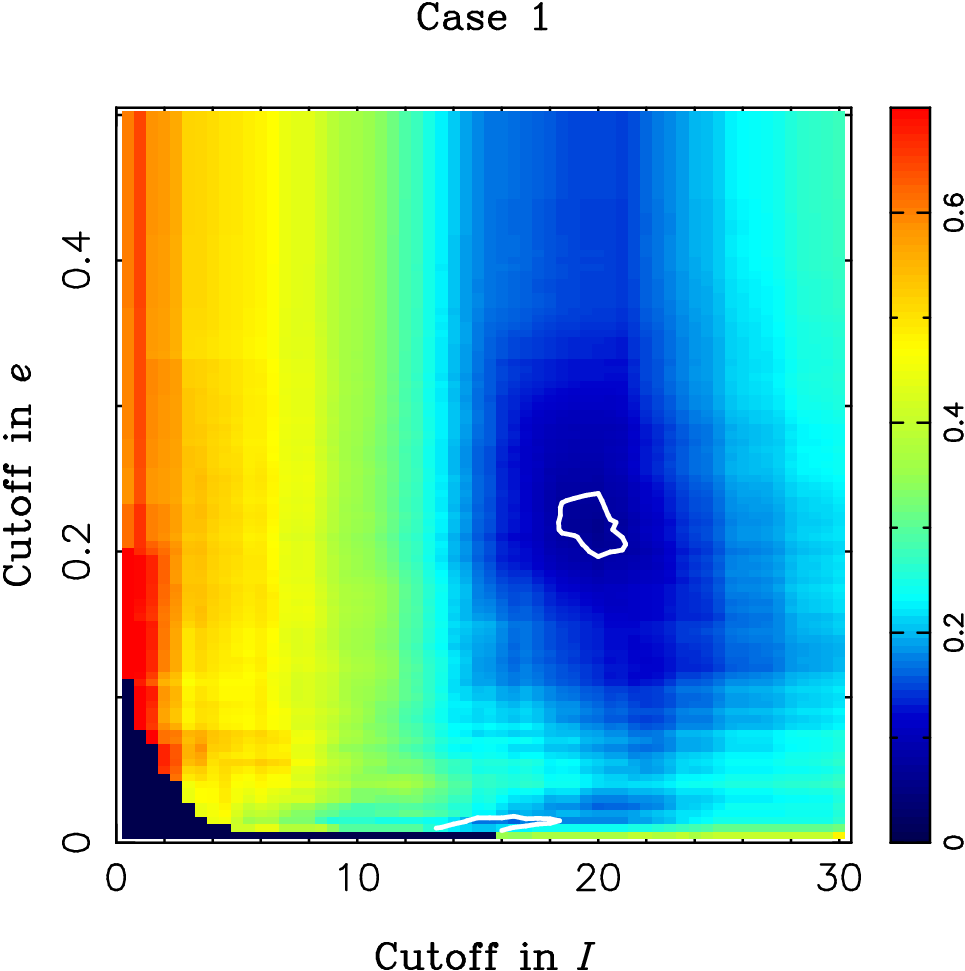}%
\includegraphics[width=0.33\textwidth]{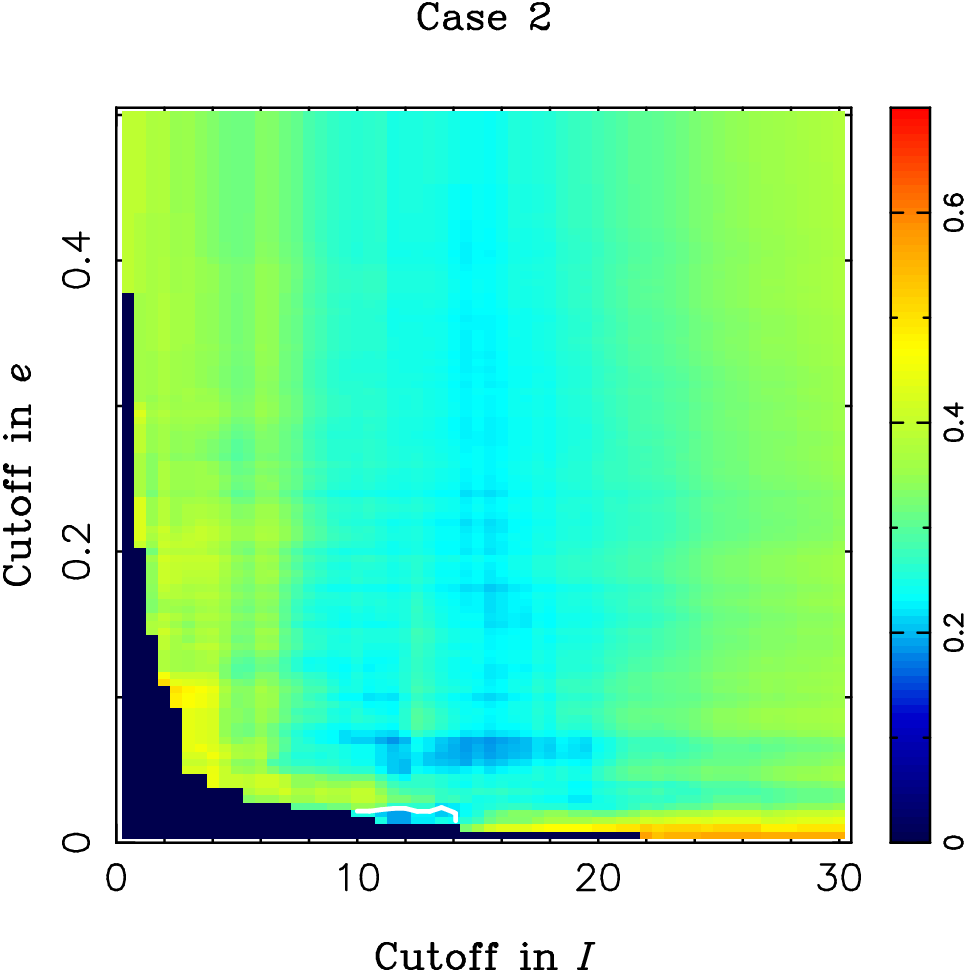}%
\includegraphics[width=0.33\textwidth]{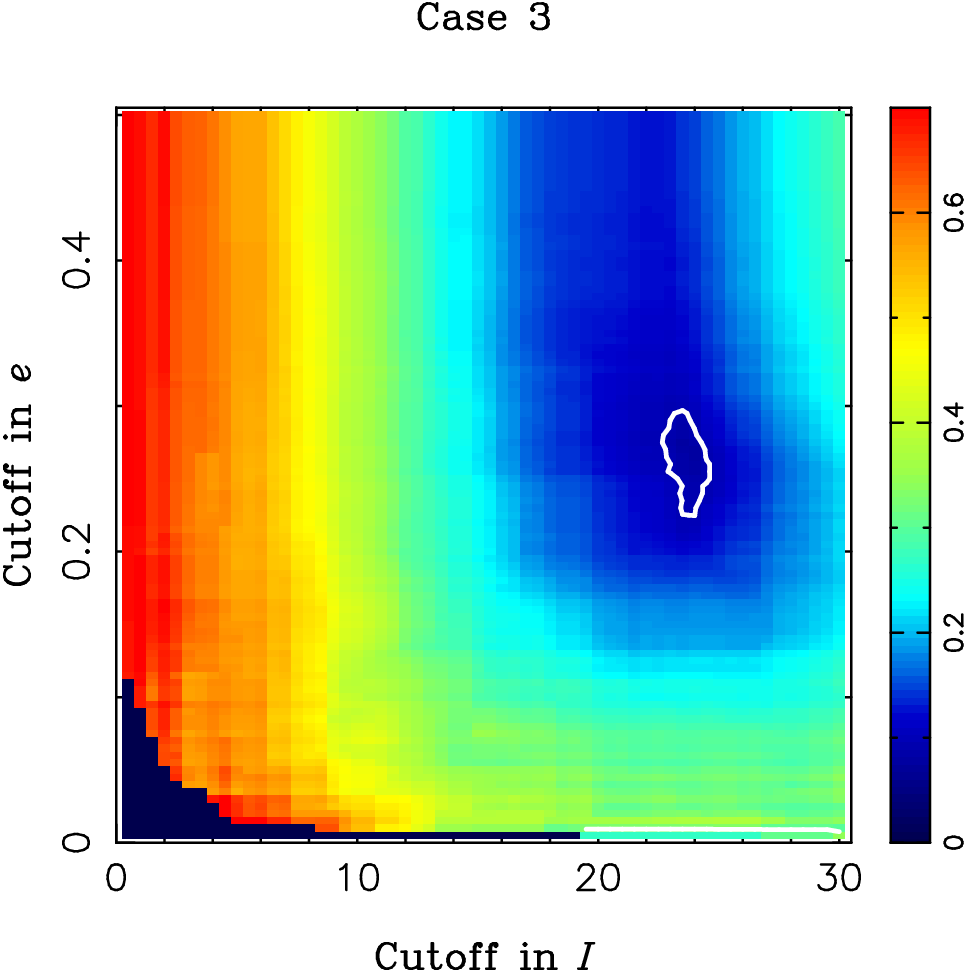}

\centering{}\medskip{}
\includegraphics[width=0.33\textwidth]{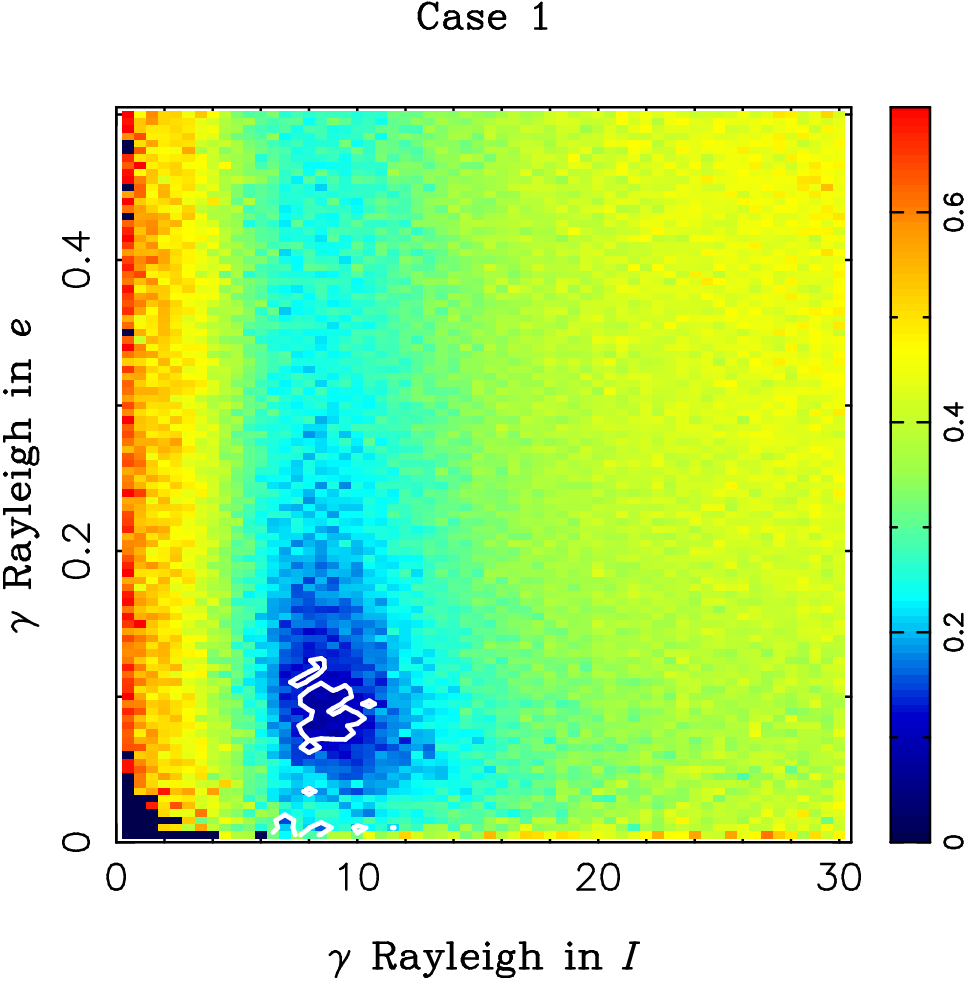}%
\includegraphics[width=0.33\textwidth]{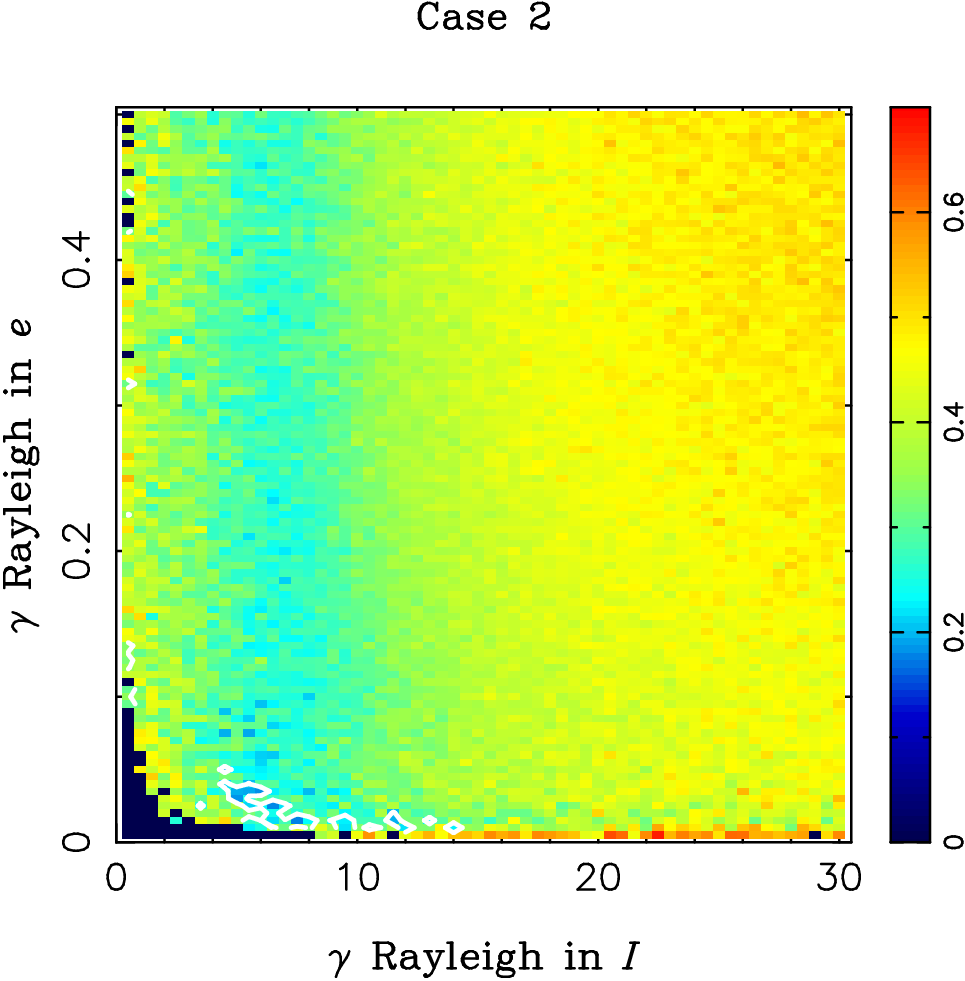}%
\includegraphics[width=0.33\textwidth]{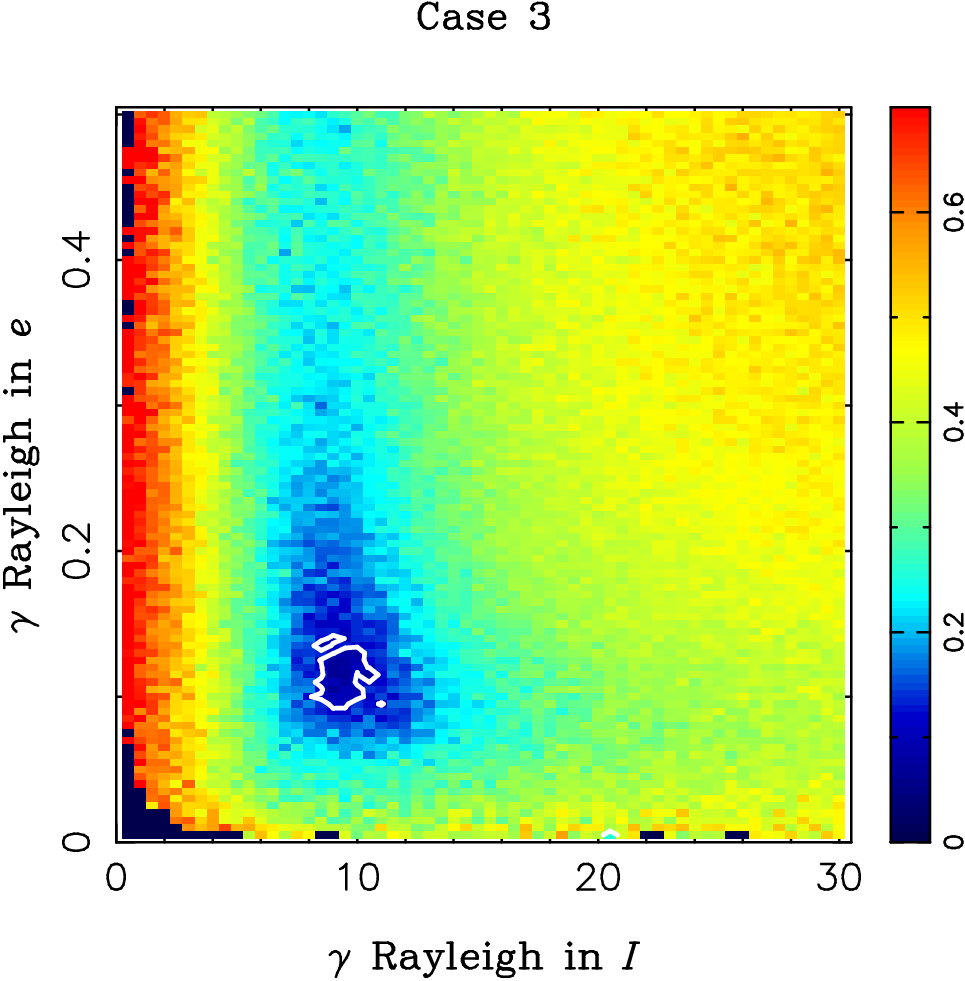}
\end{figure}

Figure \ref{ks-test-early} shows the results from the K-S test applied
to the two dimensional distributions in the $e,I$ plane, like the
one shown in Fig. \ref{comparison}c. We show examples using the initial
upper cutoff and Rayleigh distributions. Best fits with $D_{\mathrm{KS}}\leq0.1$
have been found for migration Cases 1 and 3. For migration Case 2,
it has not been possible to get a good fit because there is always
an excess of high eccentricity test particles. This indicates that
test particles in Case 2 always get too much excited in $e$, but
not necessarily in $I$. It is interesting to recall that Case 2 have
also had a bad performance in the simulations of the Galilean satellites
by \citet{2014AJ....148...25D}, because the orbits of these satellites
get too much excited. We can see that, considering upper cutoff distributions,
the primordial main belt should have been quite exited in $I$ (up
to $20^{\circ}$) and $e$ (up to 0.2). The upper limit in inclinations
is compatible with the findings by \citet{2010AJ....140.1391M}. Considering
Rayleigh distributions, the $e$ and $I$ should have been peaked
at $\sim0.1$ and $\sim10^{\circ}$, respectively. A similar result
is obtained considering Maxwell distributions. In any case, the best
fit values are consistent with the mean values of $e,I$ in the current
main belt. This result supports the idea that the presently excited
values of eccentricities and inclinations should have mostly been
acquired before the phase of planetesimal driven migration, for example,
in the Grand Tack model. The jumping Jupiter instability and the subsequent
residual migration of the outer planets help to disperse even more
the already excited $e$ and $I$, and to sculpt the main belt into
its current shape (e.g. Fig. \ref{init-fase3}).

\subsection{The Late Instability set}

\begin{figure}[H]
\protect\caption{Similar as Fig. \ref{ks-test-early} but for the LI set. The top row
corresponds to migration Case 1. The bottom row corresponds to migration
Case 3. The left column shows initial upper cutoff distributions;
middle column shows initial Rayleigh distributions, and right column
shows initial Maxwell distributions.}
\label{ks-test-late}

\centering{}\medskip{}
\includegraphics[width=0.33\textwidth]{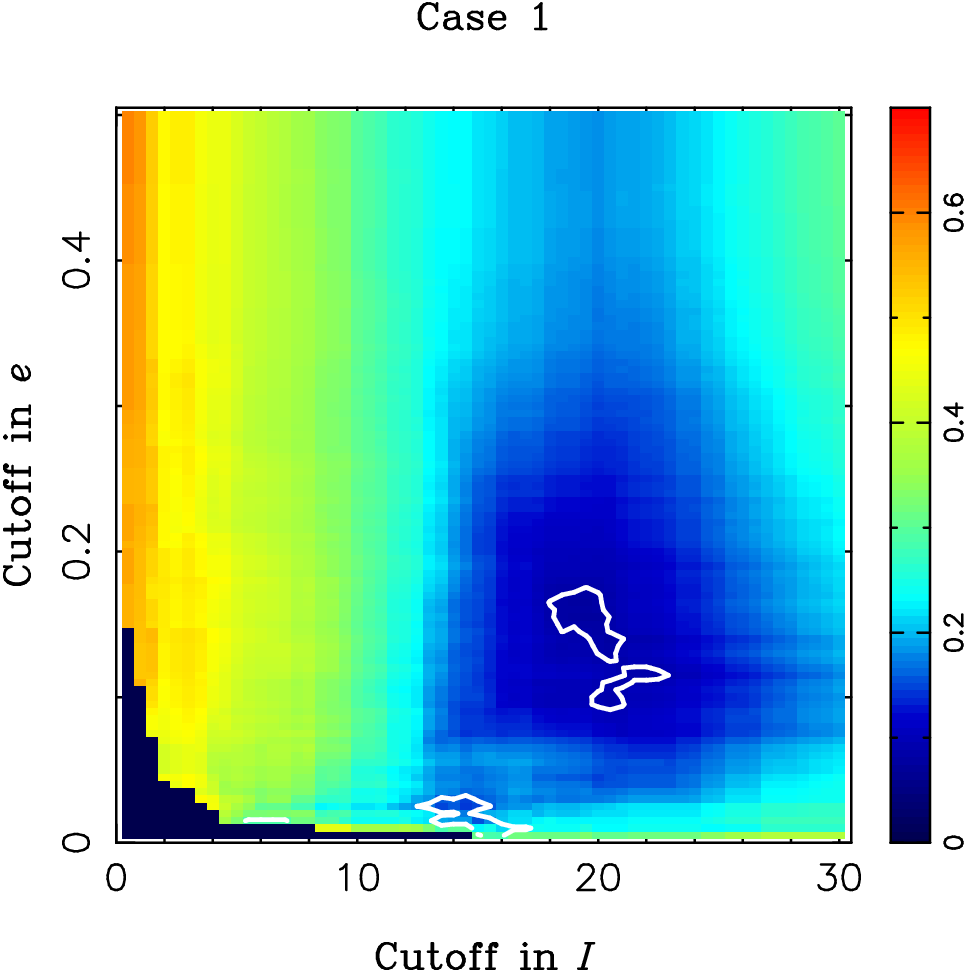}%
\includegraphics[width=0.33\textwidth]{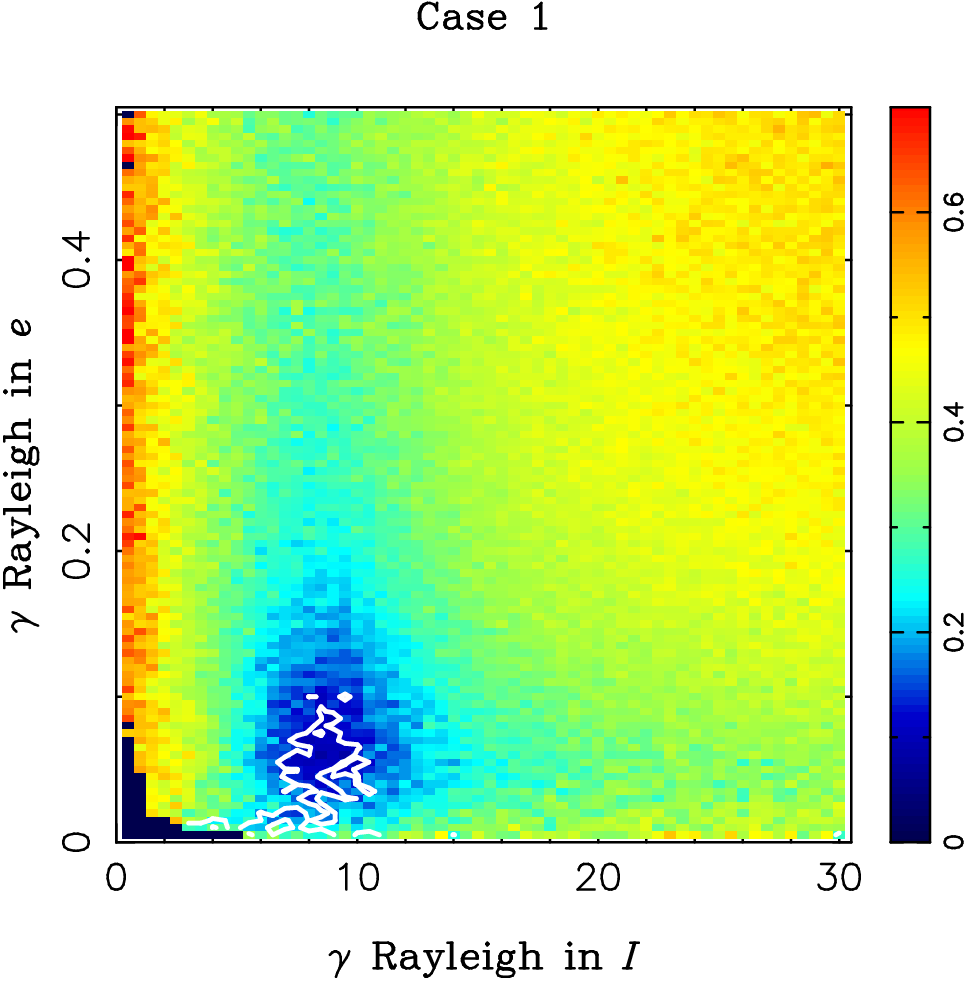}%
\includegraphics[width=0.33\textwidth]{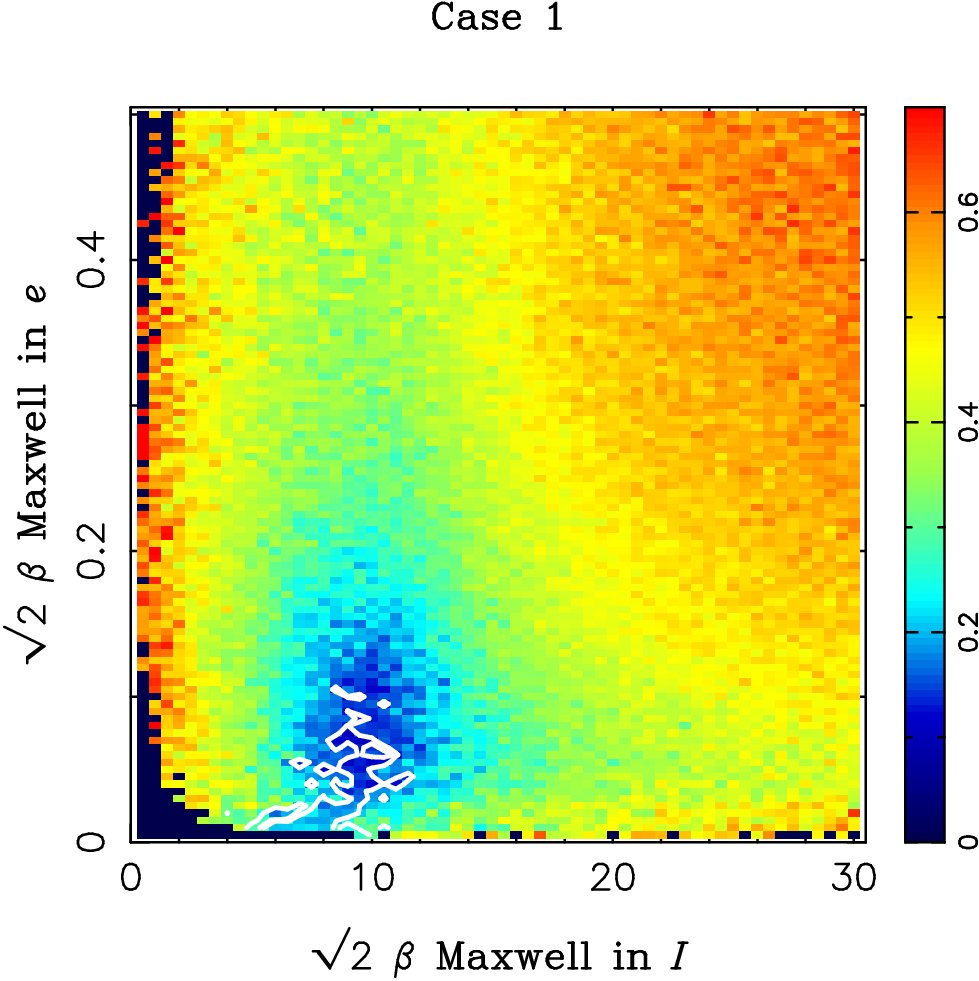}

\centering{}\medskip{}
\includegraphics[width=0.33\textwidth]{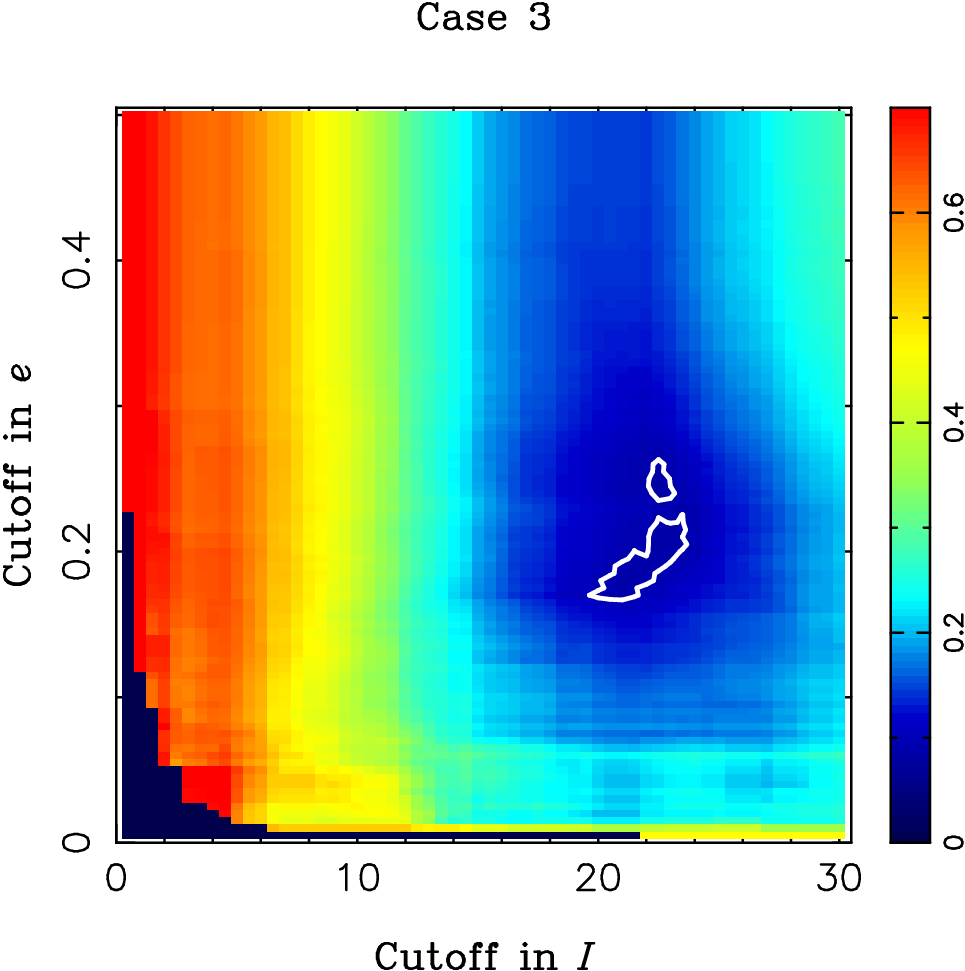}%
\includegraphics[width=0.33\textwidth]{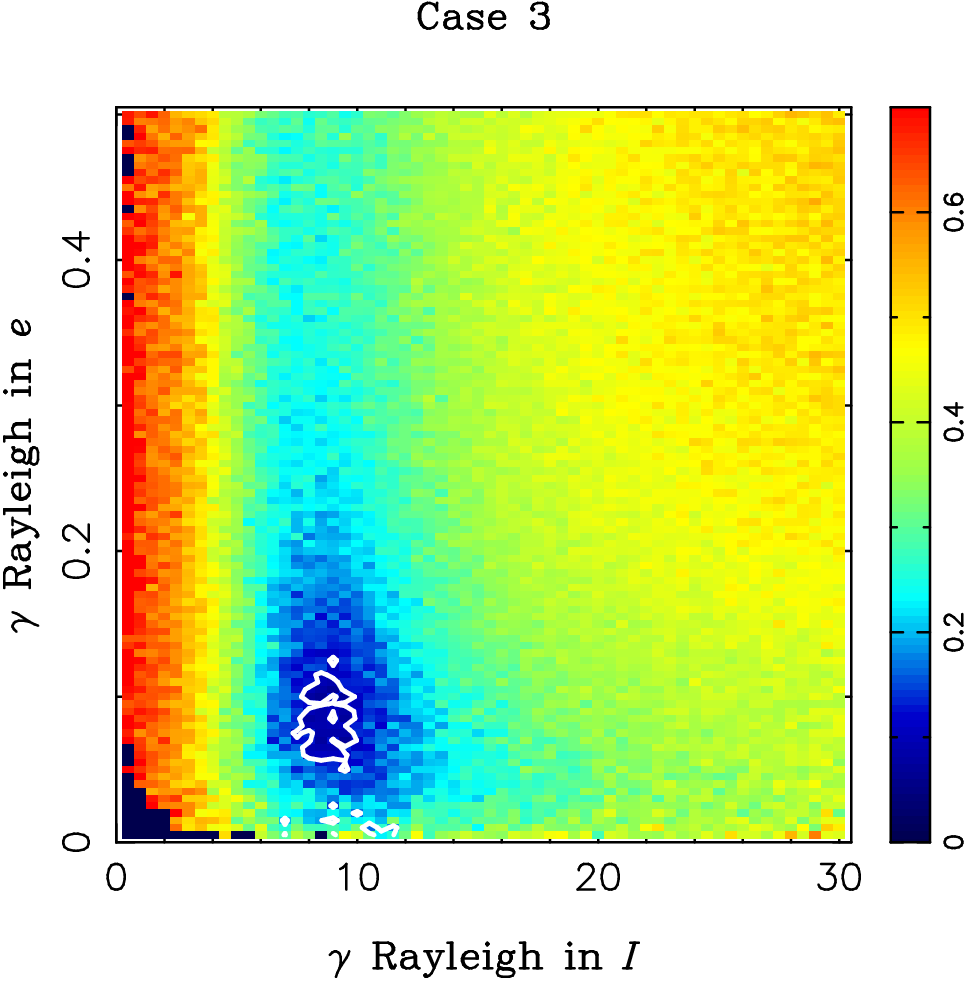}%
\includegraphics[width=0.33\textwidth]{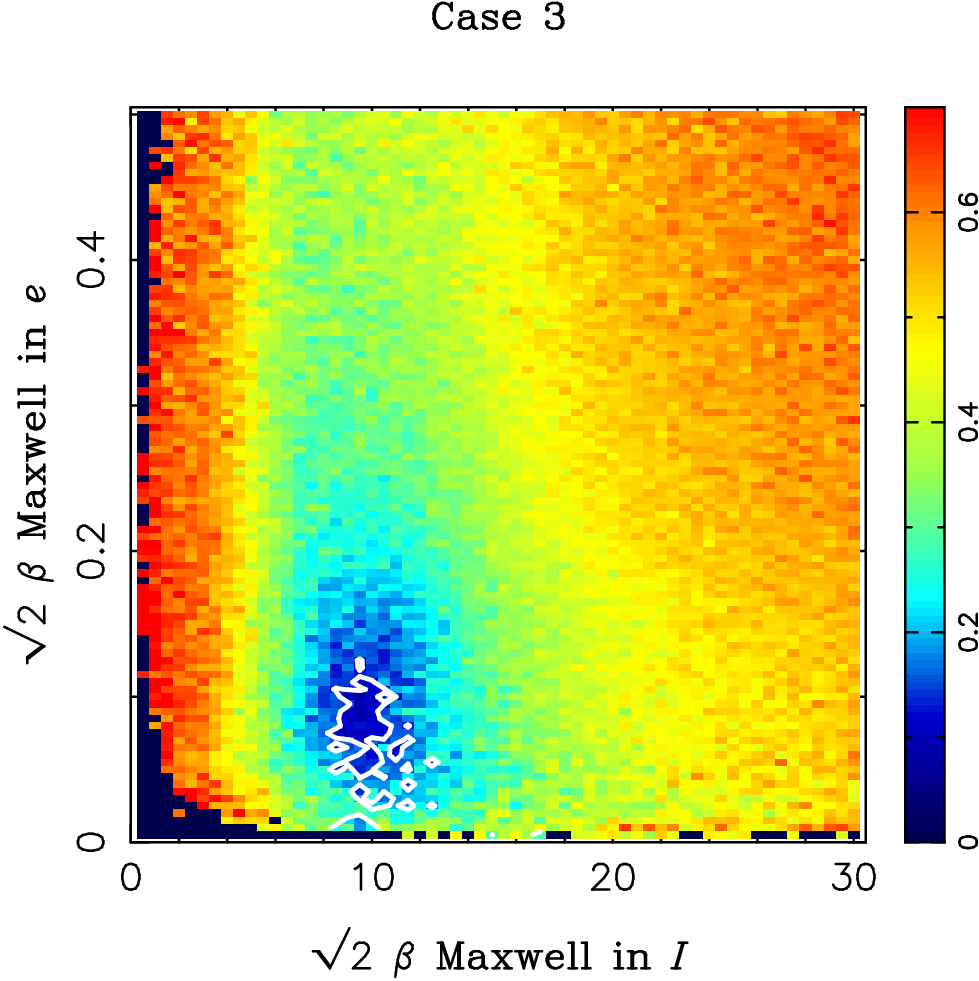}
\end{figure}

Figure \ref{ks-test-late} is similar to Fig. \ref{ks-test-early},
but for the LI set. Only results for Cases 1 and 3 are presented,
since once again Case 2 has been unable to provide a reasonable good
fit. We have not found any significant differences with respect to the
EI set, except for a slightly tighter constraint of the best fits.
This figure also shows that, in general, Maxwell distributions generate
tighter constraints than Rayleigh (also observed in the EI set). We
can conclude that, concerning the distributions in $e$ and $I$,
there is no evidence that favors the model of Early instability over
the Late one, or vice-verse. This is expected since the main difference
between the EI and LI sets concerns the stronger depletion of the
outer belt (Fig. \ref{init-fase1}), beyond the location of the 2:1
mean motion resonance, that has a small weight in the whole $e,I$
distributions.

\subsection{Implications for the Grand Tack}

The Grand Tack is a specific case of gas driven migration of the Jovian
planets. It was originally proposed to explain the low mass of Mars
and the current location of Jupiter beyond the ice line \citep{2011Natur.475..206W}.
In this model, Jupiter and Saturn formed beyond the ice line and started
to migrate inwards due to the torque exerted by the gas disk (type
II migration). Saturn migrated faster than Jupiter, and when Jupiter
had reached $\sim1.5$ au, both planets became captured and locked
in a mutual 3:2 mean motion resonance. At this point, the interaction
of the resonant configuration with the gas torque reverted the migration,
and both planets started to drift outwards \citep{2001MNRAS.320L..55M}.
By the time Jupiter reached $\sim5.5$ au, the gas had already dissipated
and the type II migration stopped. 

During the Grand Tack, Jupiter crosses the asteroid belt first inwards
and then outwards \citep{2012M&PS...47.1941W}. Therefore, this model
has deep consequences for the evolution of asteroids. In particular,
the Grand Tack is thought to be responsible for: (i) the significant
mass depletion of the asteroid belt, (ii) the primordial excitation
of the asteroids' eccentricities and inclinations, and (iii) the partial
mixing of taxonomic classes (especially the S-type and C-type) in
the main belt (for a detailed review see \citealp{2015arXiv150106204M}).

According to \citet{2012M&PS...47.1941W}, the Grand Tack would produce
an asteroidal population that can be reasonably well approximated
by a Gaussian distribution in $e$ and a Rayleigh distribution in
$I$. The best fit Gaussian distribution has mean $\mu\simeq0.38$
and standard deviation $\sigma\simeq0.17$, while the best fit Rayleigh
distribution has mode $\gamma\simeq10\text{\textdegree}$. Using these
values, we may therefore test whether the initial distributions predicted
by the jumping Jupiter instability are compatible with the expectation
for the distributions from the Grand Tack. 

\begin{figure}[H]
\protect\caption{Similar to Fig. \ref{ks-test-early}, but assuming initial distributions
as those assumed for the Grand Tack. In the left column, the abscissas
give the mode of an initial Rayleigh distribution in $I$, while the
ordinates give the standard deviation of an initial Gaussian distribution
in $e$, with fixed mean ($\mu=0.38$). In the middle column, the
abscissas give the mode of an initial Rayleigh distribution in $I$,
while the ordinates give the mean of an initial Gaussian distribution
in $e$, with fixed standard deviation ($\sigma=0.17$). Finally,
in the right column, the abscissas give the mean and the ordinates
give the standard deviation of an initial Gaussian distribution in
$e$, assuming an initial Rayleigh distribution in $I$ with fixed
mode ($\gamma=10^{\circ}$). The black triangle indicates the values
reported by \citet{2012M&PS...47.1941W}.}
\label{ks-grandtack}

\centering{}\medskip{}
\includegraphics[width=0.33\textwidth]{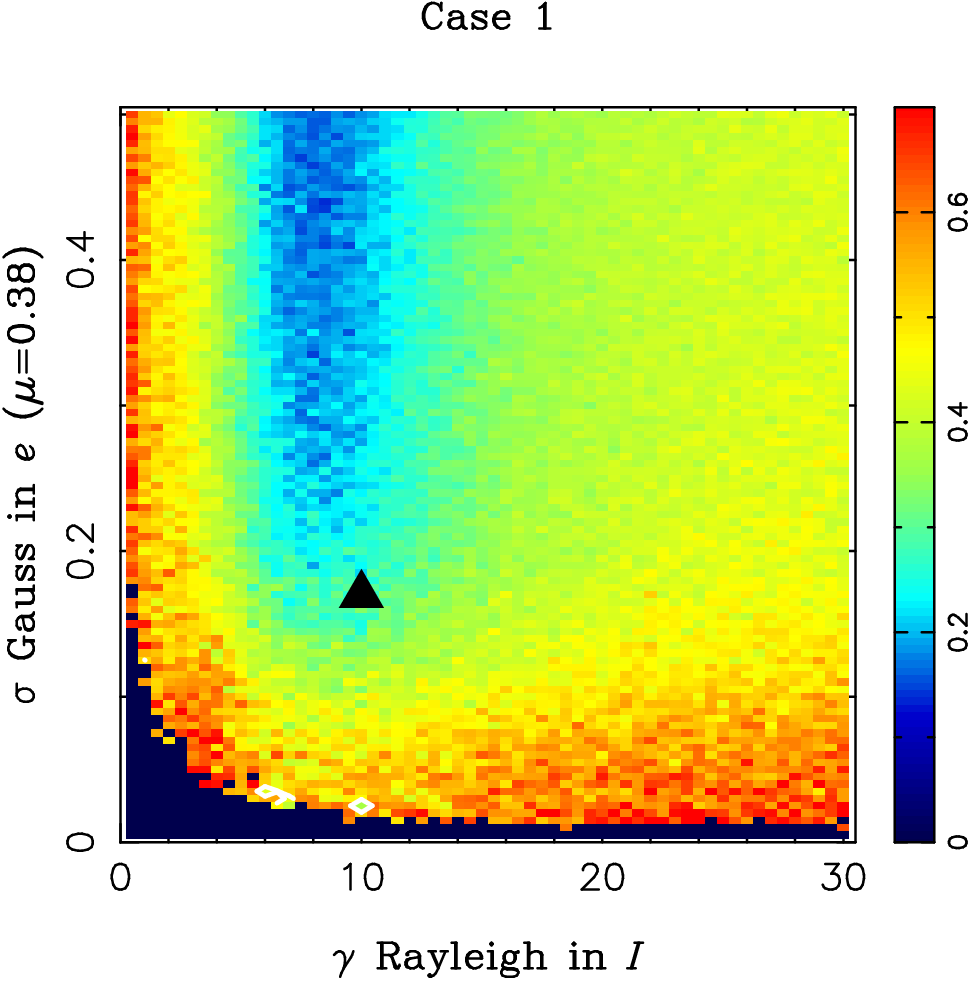}%
\includegraphics[width=0.33\textwidth]{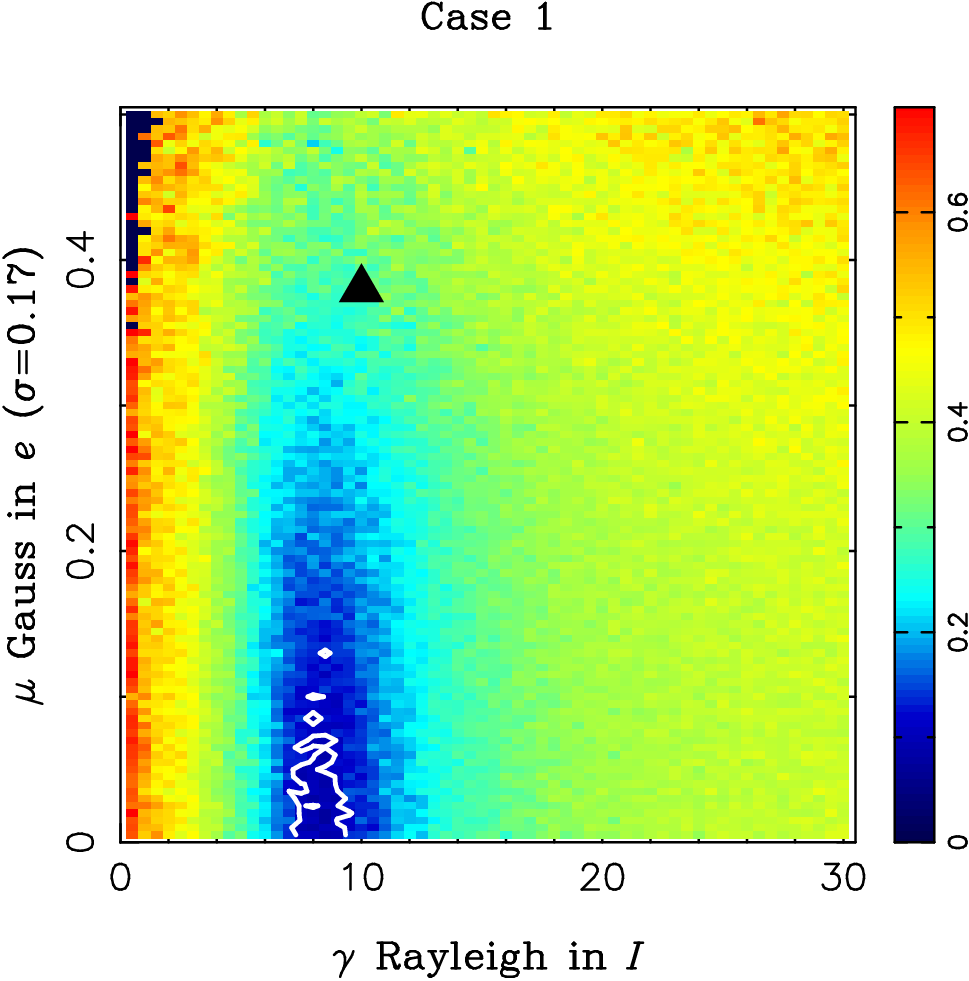}%
\includegraphics[width=0.33\textwidth]{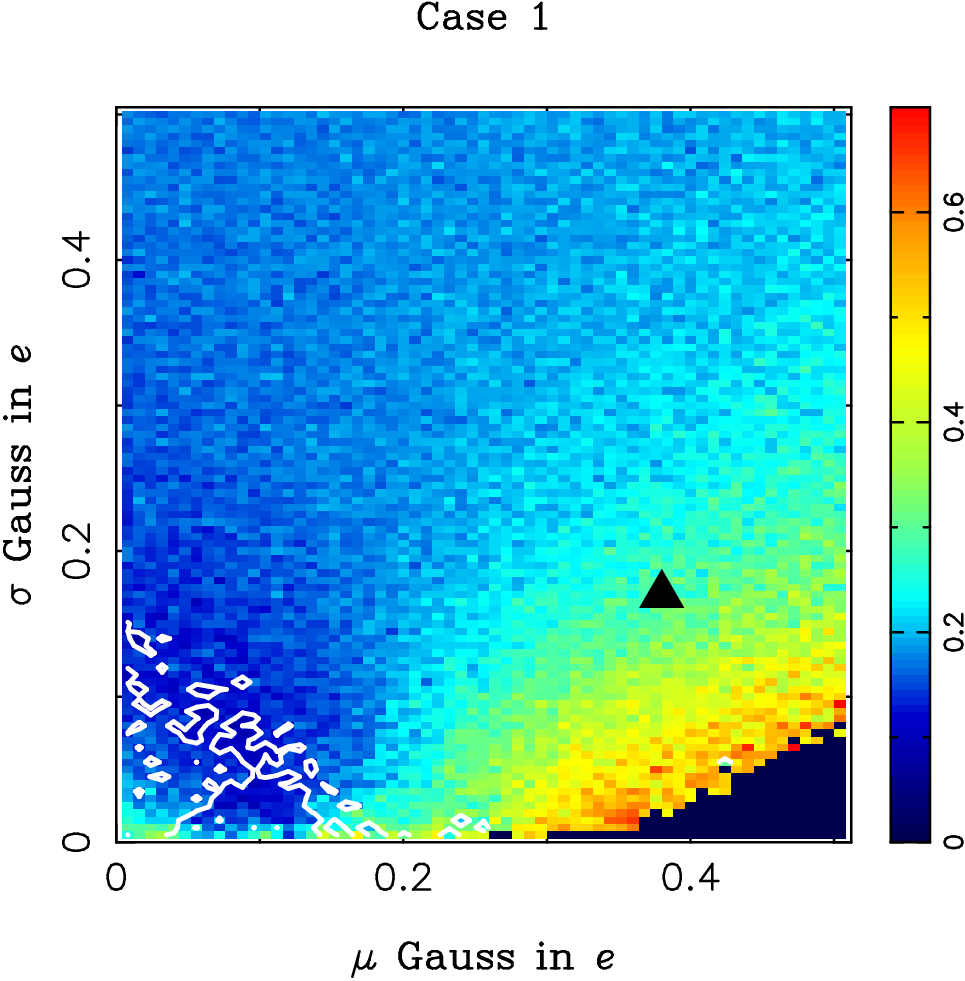}

\centering{}\medskip{}
\includegraphics[width=0.33\textwidth]{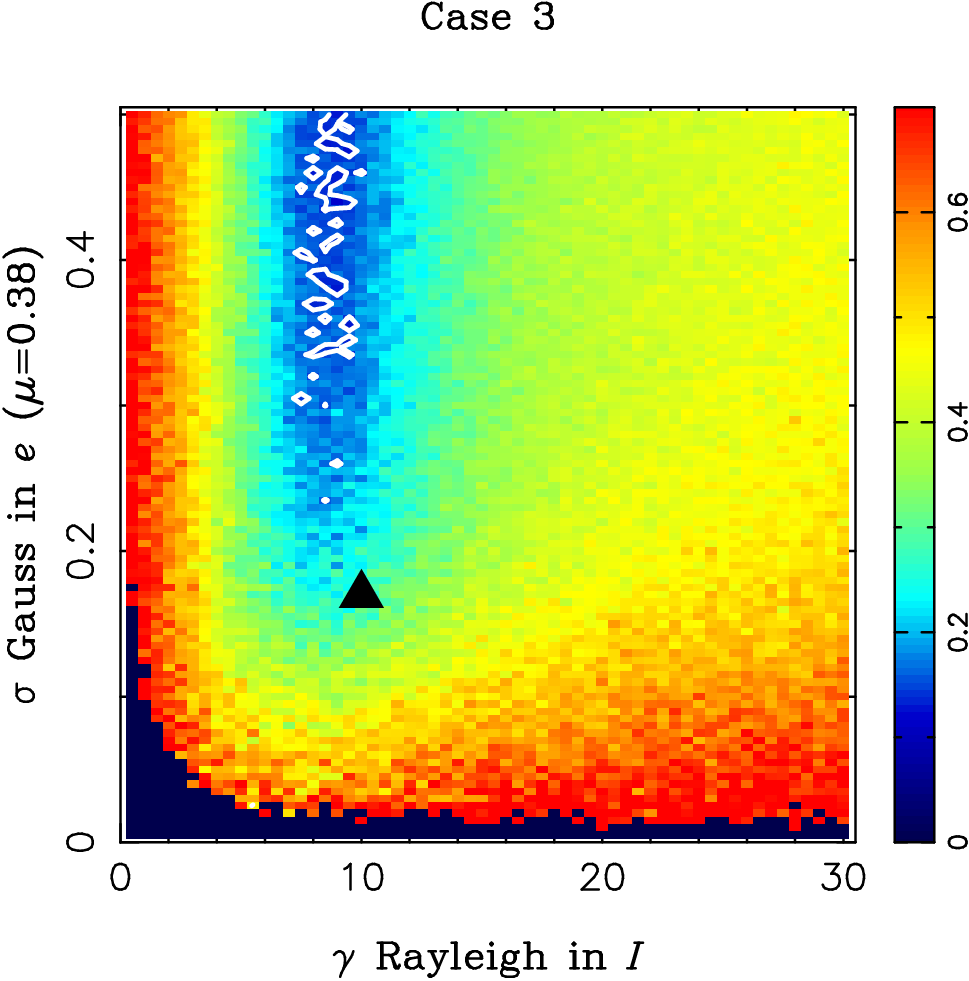}%
\includegraphics[width=0.33\textwidth]{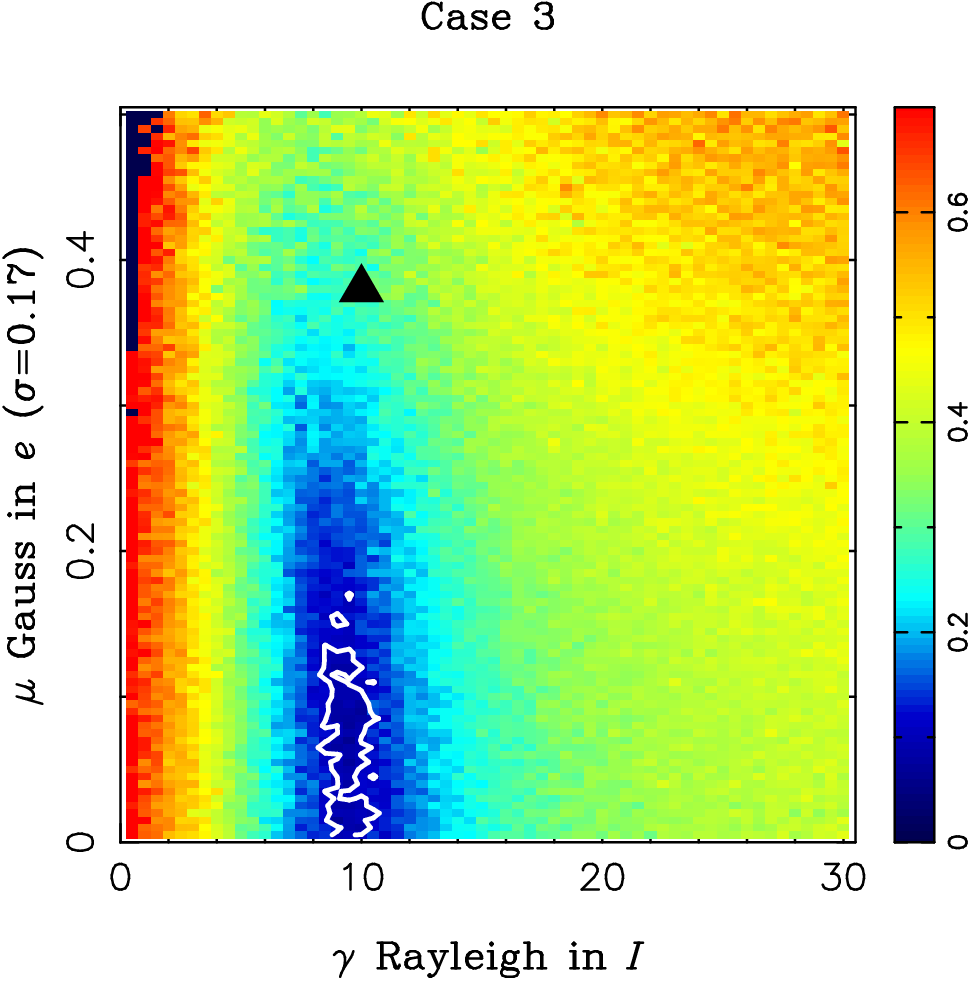}%
\includegraphics[width=0.33\textwidth]{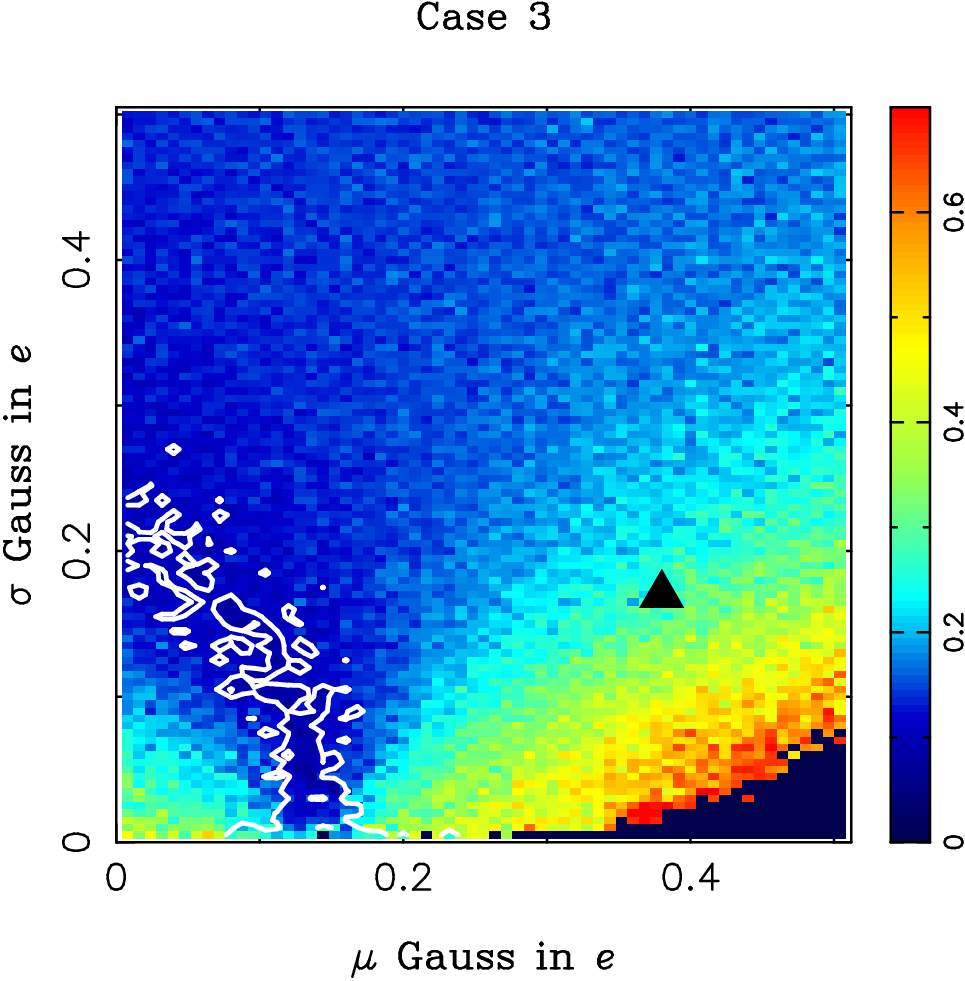}
\end{figure}

Figure \ref{ks-grandtack} shows the result. This figure is similar
to Figs. \ref{ks-test-early} and \ref{ks-test-late}, except that
the axes correspond to initial distributions like those of the Grand
Tack. This figure corresponds to the EI set; a quite similar result
has been obtained for the LI set. Once again, migration Case 2 (not
shown here) has been unable to provide a good fit due to an excess
of high eccentricity test particles in the final distribution. 

\begin{figure}[H]
\protect\caption{The same example shown in Fig. \ref{distribs} (EI set, migration
Case 1), but assuming initial distributions remapped into the nominal
Grand Tack distributions ($\mu=0.38$, $\sigma=0.17$, and $\gamma=10^{\circ}$). }
\label{distrib-gt}

\centering{}\medskip{}
\includegraphics[width=1\textwidth]{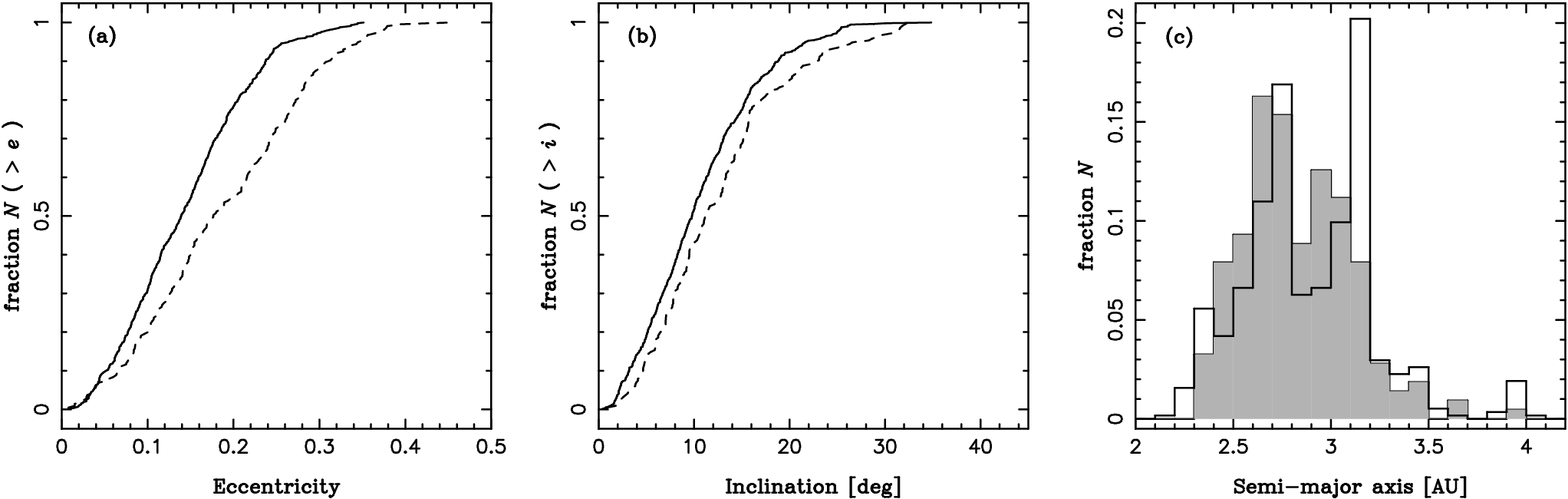}
\end{figure}

We note that, concerning the inclinations, good fits to the current
asteroid belt are obtained for a Rayleigh $\gamma\sim10^{\circ}$
that is compatible with the Grand Tack nominal value. For the eccentricities,
good fits are obtained with $\mu\sim0.1$ and $\sigma\sim0.1$-0.3.
However, if we force $\mu$ to have a large value, we need also a
large value of $\sigma>0.35$. On the other hand, if we force $\sigma$
to have a moderate value, we need a small value of $\mu<0.15$. This
result is not compatible with the nominal values derived \citet{2012M&PS...47.1941W}.
Actually, our simulations using the nominal values produce a final
distribution with a clear excess of high $e$ particles, as shown
in Fig. \ref{distrib-gt}.

There are some possible explanations for this problem with eccentricity:
\begin{itemize}
\item the distribution produced by the Grand Tack is correct, but the population
becomes significantly depleted at the high eccentricities ($e>0.2$)
by some mechanism not accounted for in our simulations. This depletion
might be caused, for example, by the terrestrial planets;
\item the distribution produced by the Grand Tack is not correct, and either
the model overestimates the stirring of eccentricities (i.e. $\mu$
is too large), or underestimates its spreading (i.e. $\sigma$ is
too small);
\item the Jovian planets evolved in a different way than considered in our
study, thus leading to different results than presented here.
\end{itemize}
An analysis of these possibilities is beyond the scope of this paper,
but the possible effect of the terrestrial planets is worth of discussion.

On one hand, the terrestrial planets might be responsible for a pre-instability
depletion of the high eccentricity asteroids, that would result in
an initial distribution peaked at low $e$ and with a moderate spreading,
as required (see Fig. \ref{ks-grandtack}, middle column). This would
put a strong constraint on the timing of the instability, because:
(i) the terrestrial planets were almost certainly not completely formed
at 10 Myr\footnote{For example, the Moon-forming impact was dated to have happened at
$\sim30$-50 Myr.} after the beginning of Phase 0, and (ii) even if they would be present
at the beginning of Phase 0, it is quite unlikely that they produce
a significant depletion of the belt in only 10 Myr of evolution. In
any case, a pre-instability depletion model seems to favor a situation
where the instability happens at later times. Actually, simulations
by Deienno et al. (2015, in preparation) indicate that an initially
cool system of terrestrial planets would require $\sim200$ Myr to
deplete the main belt so as to shift the eccentricity peak of the
Grand Tack distribution from 0.38 to $\sim0.2$ before the instability.
These authors also found that the current eccentricities and inclinations
of the asteroid belt are quite compatible with the Grand Tack initial
distribution, except at the very low values of $e$ and $I$. Their
model accounted for the terrestrial planets, but they considered a
simplified model of the jumping Jupiter instability, in which the
orbits of the major planets are artificially moved (instantaneously)
from their pre-instability configuration to their present configuration. 

On the other hand, the terrestrial planets may be responsible for
a post-instability depletion, since they will certainly continue
to erode the main belt at the high eccentricities after the instability
and until the present times. In this model, the instability may happens
at earlier times, and the main belt may still loose the necessary
fraction of high eccentricity orbits so as to reach the present distribution.

In the end, although including the terrestrial planets might help
to conciliate our results with the Grand Tack model, they would not
allow us to decide between an early \textit{vs.} late instability model.

\subsection{The distribution in semi-major axis\label{sub:The-distribution-in}}

Once we have determined the best fit parameters for the distribution
in $e$ and $I$, we may now pay attention to the distribution in
$a$. Analyzing the histograms like the one in Fig. \ref{distribs},
we conclude that our simulations have been able to reasonably reproduce
the overall distribution in $a$ of the current asteroid belt. Migration
Case 1 gives slightly better results than Cases 2 and 3. We have not
found significant differences between the EI and LI sets. We also
realized that the distribution in $a$ is nearly independent on the
assumed initial distributions in $e$ and $I$. 

A major discrepancy between our simulations and the real asteroid
belt occurs in the interval $2.8<a<3.0$ au. Today, this region of
the belt shows a much lower density of asteroids compared to the neighboring
regions. The reason for this is still unclear (e.g. \citealp{2009Natur.457.1109M}).
\citet{2013A&A...551A.117B} refer to this region as the ``pristine
zone''. In our simulations, we have not been able to reproduce the
low density of the pristine zone. Actually, in some simulations the
density we have obtained can be up to twice the observed one. 

\begin{figure}[H]
\protect\caption{Results of the $\chi^{2}$ test (color scale) for the semi-major axis
distribution of the EI set in the three migration cases considered
here. The horizontal and vertical axes give the values of the parameters
$a_{c}$ and $\varepsilon$ of the band cut initial distribution in
$a$. A cut level $\varepsilon=0$ means that all the test particles
within a band centered at $a_{c}$ have been removed. A cut level
$\varepsilon=1$ means that no test particle has been removed. Good
fits are obtained for $\chi^{2}\leq0.14$ (note that $\chi^{2}$ is
always larger than 0.1). Initially, $e$ and $I$ were assumed to
follow the Rayleigh distributions with modes $\gamma=0.1$ and $10^{\circ}$,
respectively. The black rectangle shown in migration Case 1 indicates
a region where the fits are the best. Similar results were obtained
for the LI set.}
\label{chi2-test}

\centering{}\medskip{}
\includegraphics[width=0.33\textwidth]{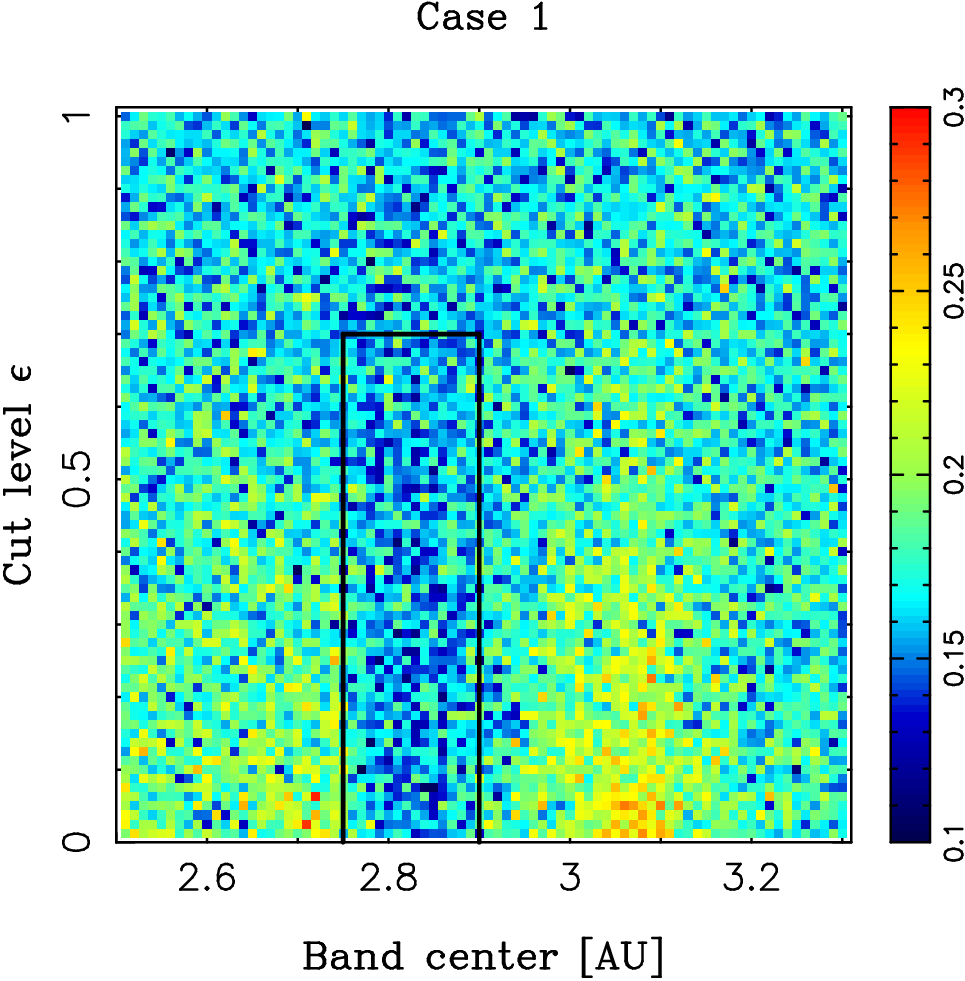}%
\includegraphics[width=0.33\textwidth]{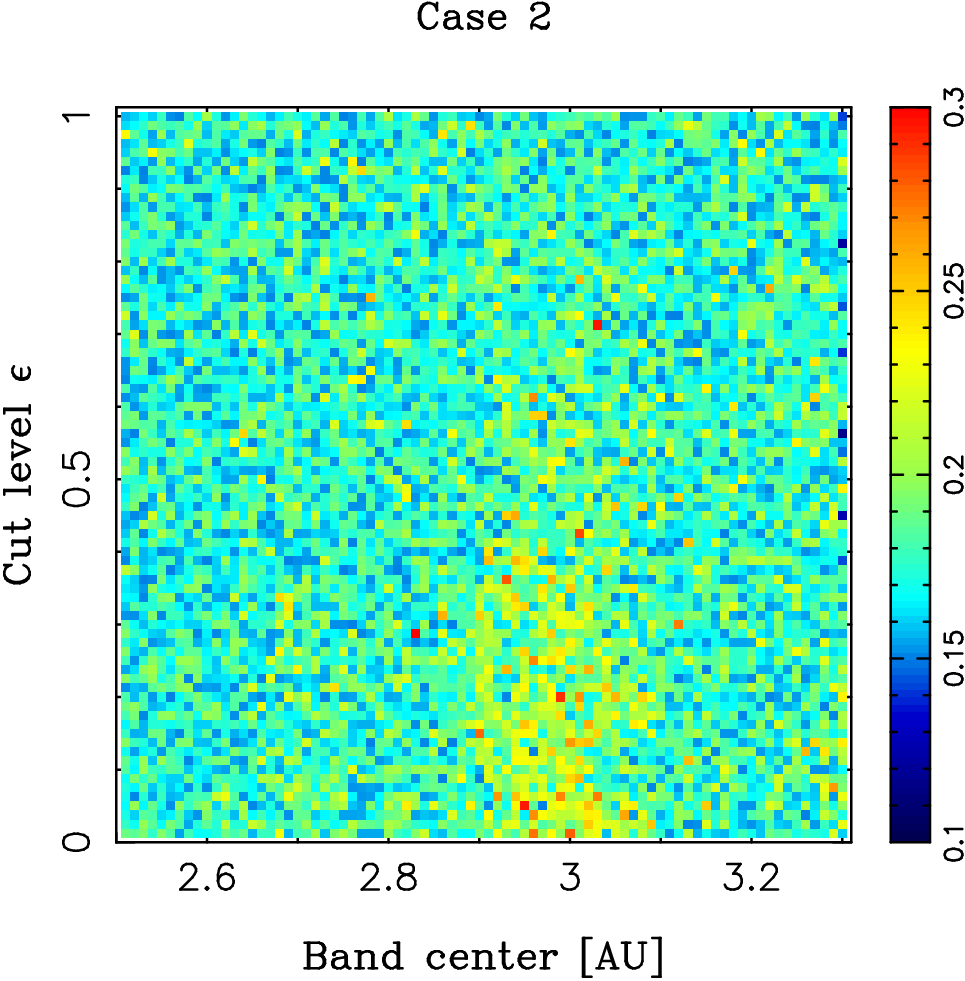}%
\includegraphics[width=0.33\textwidth]{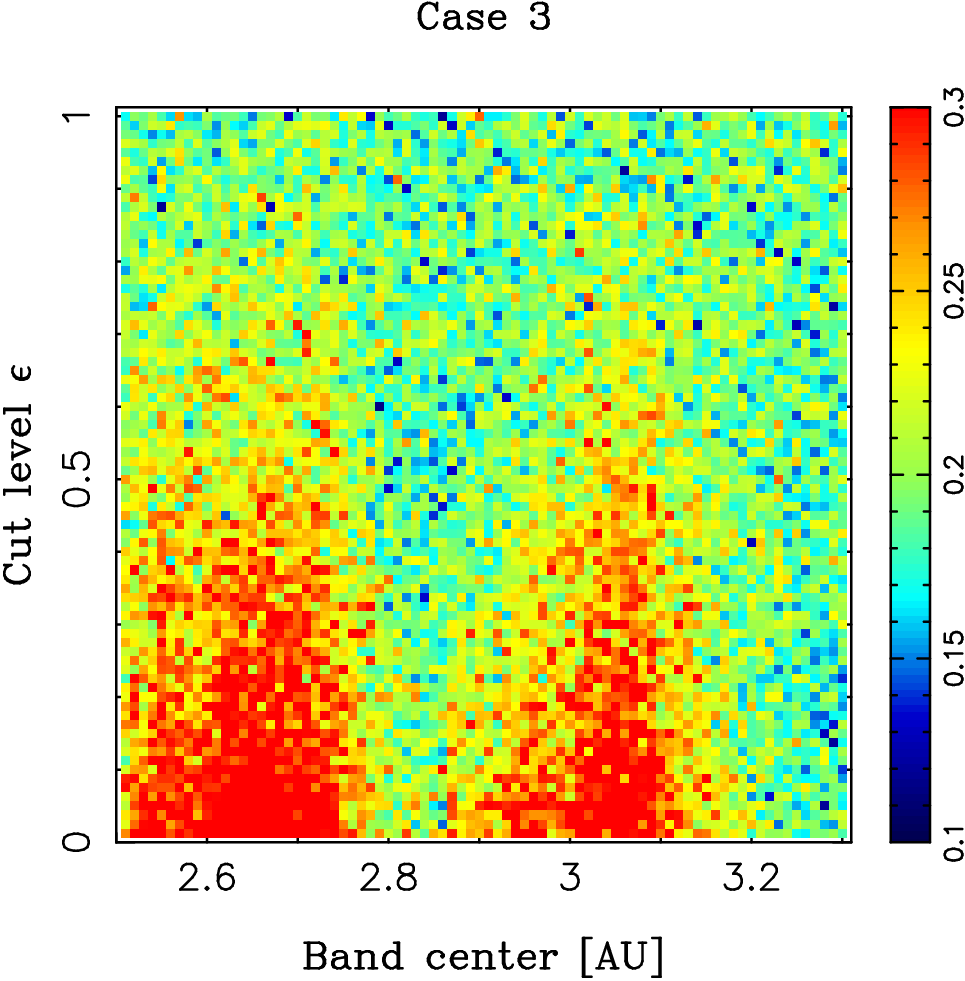}
\end{figure}

Here, we test whether a primordial depletion of this zone (i.e. before
the jumping Jupiter instability) could explain the current density.
We do not intend to provide a dynamical explanation for such hypothetical
depletion, but simply to determine if it could be a plausible alternative.
The depletion has been simulated by remapping the initial density
distribution in $a$ into a ``square band cut'' distribution, i.e.:
\begin{equation}
p(a)=\left\{ \begin{array}{l}
1\qquad\mathrm{if}\;a<a_{\mathrm{min}}\;\mathrm{or}\;a>a_{\mathrm{max}}\\
\varepsilon\qquad\mathrm{if}\;a_{\mathrm{min}}\leq a\leq a_{\mathrm{max}}
\end{array}\right.\label{eq:sma}
\end{equation}
with $0\leq\varepsilon\leq1$ ($\varepsilon=0$ means that all the
test particles within the band are removed). The final density distribution
obtained from the remapped initial distribution has been fitted to
the real asteroid belt applying a $\chi^{2}$ test. We have considered
a band cut distribution with a fixed band width $a_{\mathrm{max}}-a_{\mathrm{min}}=0.15$
au, compatible with the current width of the pristine zone. The band
center $a_{\mathrm{c}}$ and the cut level $\varepsilon$ have been
taken as free parameters of the fit. Figure \ref{chi2-test} shows
the results for the EI set in the three migration cases. The color
scale gives the values of $\chi^{2}$, and we have considered that
good fits are obtained for $\chi^{2}\leq0.14$ (dark blue). In these
examples, we have assumed that $e$ and $I$ initially followed the
Rayleigh distributions with $\gamma=0.1$ and $10^{\circ}$, respectively,
corresponding to the best fits for these orbital elements. 

\begin{figure}[H]
\protect\caption{The same example shown in Fig. \ref{distribs} (EI set, migration
Case 1), but now assuming initial distributions remapped into the
Rayleigh distributions in $e$ and $I$. The modes in $e$ and $I$
are $\gamma=0.1$ and $10^{\circ}$, respectively. No remapping of
the distribution in $a$ has been applied. Note the larger depletion
in the ``pristine zone'' (2.8 to 3.1 au) compared to Fig. \ref{distribs}.}
\label{distrib-remap}

\centering{}\medskip{}
\includegraphics[width=1\textwidth]{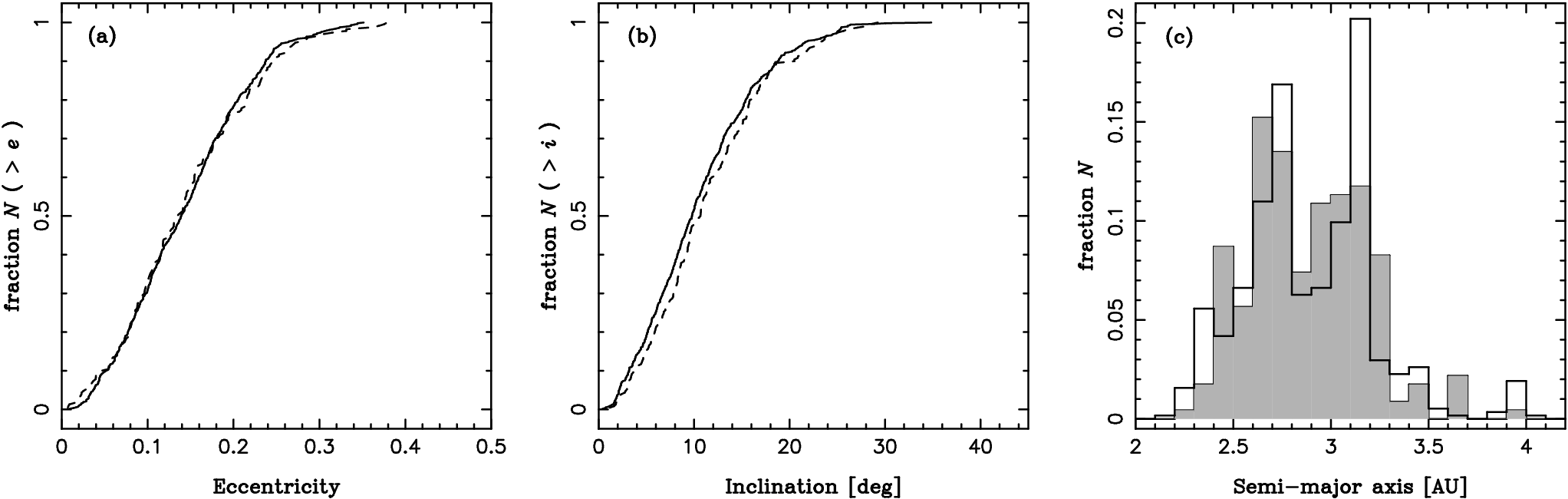}
\end{figure}

In Case 1, we observe that $\varepsilon\lesssim0.3$ (i.e. $\apprge30$\%
of depletion) around 2.85 au (black rectangle) appears to provide
slightly better fits to the $a$ distribution. However, the results
are inconclusive and, in principle, good fits could also be obtained
without any depletion at all (i.e. $\varepsilon=1$). An example is
shown in Fig. \ref{distrib-remap}. This is the same simulation shown
in Fig. \ref{distribs}, but remapping the initial distributions in
$e$ and $I$ only. Apart from the very good matches in the $e$ and
$I$ distributions, as expected, we see that the distribution in $a$
shows a better match in the pristine zone, even without having applied
any remapping to its initial distribution. On the other hand, in Case
2 no good fit can be found in spite of the application of a band cut
distribution in $a$. Finally, in Case 3, an initial band cut distribution
in $a$ produces a much worse fitting than the original distribution.
The reason for these differences among the three cases may be related
to the amount of dispersion in $a$ caused in each migration model.
Table \ref{mean-dispersion} shows the median, mean and maximum dispersion
in orbital elements obtained in the different cases. These values
are reflecting the behavior of the fifth planet in each simulation,
as discussed in Sect. \ref{sub:Phase-1} (Brasil et al., 2015, in
preparation). It appears that Case 2 causes too much dispersion in
$a$ as to blur any structure in the initial distribution. On the
other hand, Case 3 does not disperse enough the semi-major axes, and
any sharp structure in the initial distribution will remain in the
final distribution. For example, applying a large initial depletion
around 2.6 au leaves this region almost depleted over the whole simulation
(because it is not replenished by particles dispersed from the neighboring
regions), thus producing the large $\chi^{2}>0.25$ values observed
in Fig. \ref{chi2-test}.

The values of $\delta a$ shown in Table \ref{mean-dispersion} also
allow us to infer that the dispersion caused by the jumping Jupiter
instability would not be enough to produce a significant mixing of
the taxonomic classes in the main belt, as suggested by the observational
evidence (\citealp{1982Sci...216.1405G}; \citealp{2013Icar..226..723D},
\citeyear{2014Natur.505..629D}). This supports the idea that the
mixing of taxonomies is the consequence of a pre-instability mechanism,
as for example, the Grand Tack model.

\subsection{The Hilda and Trojan populations\label{sub:The-Hilda-and}}

As discussed in Sect. \ref{sub:Phase-0}, any primordial population
of Hildas and Trojans that might have existed after the Grand Tack,
is strongly depleted already during Phase 0. Even in the most optimistic
model of an Early Instability, the HG is reduced to $\sim10$\%, and
the TS to less than 1\% (Table \ref{statistic-of-surviving}). We
have also verified that these remnant populations do not survive the
jumping Jupiter instability. Of the initially 15~000 test particles
in the HG and TS populations at the beginning of Phase 1 (red dots
in Fig. \ref{init-fase1}), less than 1\% have survived the instability
but all of them have been scattered out of the 3:2 and 1:1 mean motion
resonances with Jupiter. We conclude that in an Early Instability
hypothesis, the survival probability of a post Grand Tack population
is $<10^{-5}$ for the HG and $<10^{-6}$ for the TS. These probabilities
reduce further by two orders of magnitude in a Late Instability hypothesis.

\begin{figure}[H]
\protect\caption{\emph{Top row:} Distribution of the MB test particles that became
implanted in the Hilda region (blue dots) at the end of Phase 1. Gray
dots represent the MB population. Red dots are the implanted test
particles that survived until the end of Phase 3. \emph{Bottom row:}
Initial distribution, at the beginning of Phase 1, of the MB test
particles that become implanted in the Hilda region. Gray dots represent
the whole MB initial population. This example corresponds to the EI
set, migration Case 3. }
\label{hilda-fase1}

\centering{}\medskip{}
\includegraphics[width=1\textwidth]{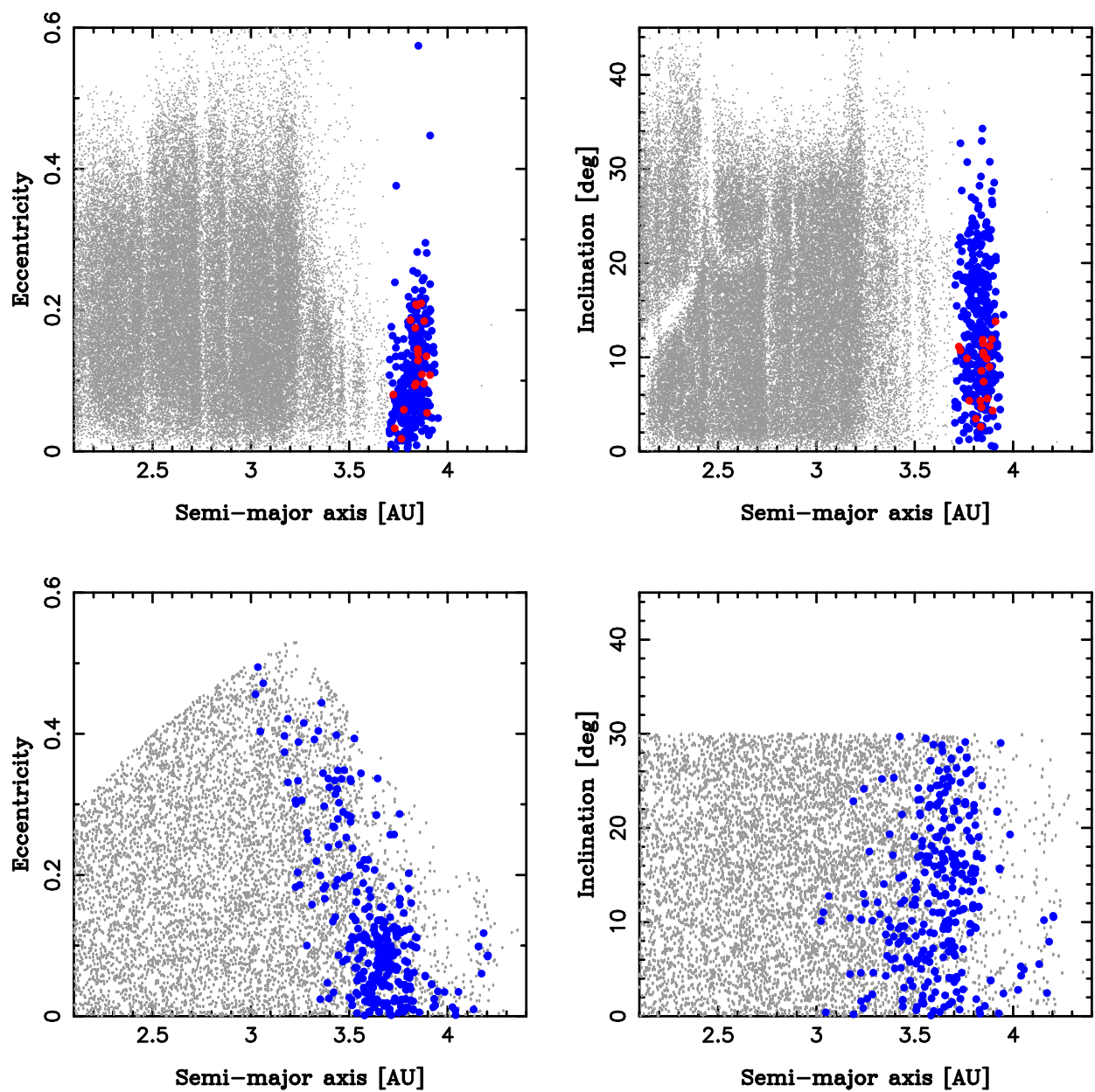}
\end{figure}

On the other hand, we have realized that after the instability a number
of MB test particles have became apparently implanted in the region
of the HG and TS populations. An example is shown in Fig. \ref{hilda-fase1},
top row, where the blue dots correspond to the test particles that
survived in the HG and TS regions at the end of Phase 1. We have also
verified that most of these implanted test particles had initial orbits
at the beginning of Phase 1 with $a>3$ au (Fig. \ref{hilda-fase1},
bottom row). Since Jupiter is at $\sim5.5$ au before the instability,
this source region is beyond the 5:2 mean motion resonance with Jupiter,
i.e. it is equivalent to the current outer main belt. Finally, our
simulations also indicate that less than 7\% of the implanted populations
have survived the phase of residual migration and are still active
at the end of Phase 3 (red dots in Fig. \ref{hilda-fase1}). This
allows us to make an estimation of the implanting probability, simply
dividing the number of surviving test particles at the end of Phase
3 by the total number of test particles with $a>3$ au at the beginning
of Phase 1 (it is worth recalling that the implanted test particles
have not been down-sampled in the transition from Phase 2 to Phase
3). The estimated probability for implanting HG particles is $\sim4\times10^{-4}$
in the most optimistic case (EI set, migration Case 3) and $<4\times10^{-5}$
in the most pessimistic case (LI set, migration Case 1). For implanting
TS particles, the probability is $<10^{-5}$-$10^{-6}$.

In order to estimate the current fraction of Hilda asteroids that
could have been implanted from the MB, we have proceeded as follows:
\begin{enumerate}
\item we have verified that, in general, our simulations reproduce quite
well the distribution of asteroids in the interval $3.2\leq a\leq3.5$
au (see for example Figs. \ref{distribs}c, \ref{distrib-gt}c and
\ref{distrib-remap}c), at least in terms of relative density. This
region is presently occupied by the group of Cybele, and we will refer
to it as the Cybele region;
\item we have also verified that the test particles that have survived in
the Cybele region have originated in the same region as the implanted
Hildas, i.e. our simulated Cybeles had $a>3$ au at the beginning
of Phase 1;
\item if $n_{C}$ is the number of Cybeles and $n_{H}$ is the number of
Hildas that have survived in our simulations at the end of Phase 3,
then the present fraction of possibly implanted Hildas can be computed
as:
\begin{equation}
f_{\mathrm{imp}}=\frac{n_{H}}{f_{\mathrm{down}}n_{C}}\frac{N_{C}}{N_{H}}\label{eq:fimp}
\end{equation}
where $f_{\mathrm{down}}$ is the down sampling factor that has been
applied to the Cybele region, $N_{C}$ is the current number of asteroids
in the Cybele region, and $N_{H}$ is the current number of Hilda
asteroids (both with $H<11$, corresponding to diameter greater than
$\sim40$ km for albedo $\sim0.05$).
\end{enumerate}
We computed that $f_{\mathrm{imp}}<5\%$ in the simulations with the
EI set, and $f_{\mathrm{imp}}<1\%$ in the simulations with the LI
set. For the Trojans, these fractions are two orders of magnitude
smaller. This result has implications for the current distribution
of taxonomical classes among the Hildas. Assuming that the source
of the implanted Hildas is dominated by C-type asteroids, as is the
case of the current outer main belt, we could expect to find today
1 or 2 implanted C-type Hildas with $H<11$. Actually, up to this
magnitude limit, the Hilda group is dominated by P-type and D-type
asteroids (\citealp{1995A&A...302..907D}; \citealp{1997A&A...323..606D})\footnote{See also \url{http://ssd.jpl.nasa.gov/sbdb_query.cgi}},
and only one C-type asteroid is known: (334) Chicago. 

The above results support the idea that current Hildas and Trojans
are not primordial, but were captured from other Solar System populations
during the migration instability. \citet{2005Natur.435..462M} and
\citet{2013ApJ...768...45N} have shown that Trojans can be captured
from the disk of planetesimals initially located beyond the orbit
of Neptune. It is possible that Hilda asteroids could also be captured
from this population \citep{2009Natur.460..364L}, but, at variance
with Trojans, a small fraction ($<5$\%) of them could have also been
captured from the outer asteroid main belt.

\section{Conclusions\label{sec:Conclusions}}

In this work, we have investigated the evolution of the asteroid belt
during the jumping Jupiter instability related to the planetesimal
driven migration of the Jovian planets. We have simulated the dynamical
behavior of test particles from the epoch before the instability to
the present days. We have considered sets of initial conditions distributed
uniformly, that have been remapped to different non uniform distributions
in order to find a best match between the final results and the current
asteroid belt. Our conclusions can be summarized as follows:
\begin{itemize}
\item we have not found any indication that favors a model in which the
instability occurs early (soon after the end of the gas driven migration),
with respect to a model in which the instability occurs later (at
a time compatible with the Late Heavy Bombardment epoch);
\item of the three models of jumping Jupiter instability that we have tested,
the one identified as Case 2 has been unable to produce any good fit
to the present asteroid belt, because the final eccentricities get
too much excited;
\item in the other two models (Cases 1 and 3), the best fits indicate that,
prior to the instability, the asteroid belt should have been quite
excited in eccentricities (0.1 to 0.2) and inclinations ($10^{\circ}$
to $20^{\circ}$). The main effect of the jumping Jupiter instability
is only to cause a moderate dispersion of the $e,I$ values. This
supports the idea that most of the current excitation of the main
belt occurred before the instability, as it would be the case of the
Grand Tack model; 
\item concerning the upper limit for the initial inclination distribution,
we have found the same results as \citet{2010AJ....140.1391M};
\item our results predict that before the instability, the asteroid belt
should have had a Rayleigh (or Maxwell) distribution in inclinations
peaked at $\sim10^{\circ}$. This is compatible with the value found
by \citet{2012M&PS...47.1941W} for the distribution resulting from
the Grand Tack model;
\item on the other hand, our results predict that the eccentricities of
the pre-instability asteroid belt should have been peaked at $\sim0.1$,
which is smaller than the value of 0.38 assumed by \citet{2012M&PS...47.1941W}.
This discrepancy could be resolved invoking an additional mechanism
that would deplete the high $e$ orbits during the transition between
the Grand Tack and the jumping Jupiter instability. In principle,
the terrestrial planets, which have not been considered in our simulations,
could account for such depletion;
\item our model is able to reproduce quite well the density distribution
in semi-major axis of the main belt up to 3.5 au, except in the the
interval between 2.8 and 3.1 au, where we have only been able to partially
reproduce the very small density observed today. This could be indicating
that such region became even more depleted by some other (not yet
understood) mechanism after the jumping Jupiter instability;
\item the median and mean dispersion in semi-major axis caused by the instability
does not seem to be enough to provide the required mixing of taxonomic
classes in the main belt;
\item the jumping Jupiter instability is able to implant asteroids from
the outer main belt into the Hilda group with a very low probability,
although enough to explain the presence of a few big Hildas belonging
to the C taxonomic class.
\end{itemize}

\acknowledgements{We wish to thank the anonymous referee for his helpful
comments and suggestions. This work has been supported by the Brazilian 
National Research Council
(CNPq) through fellowship 312292/2013-9 and grant 401905/2013-6
within the Science Without Borders Program, and by NASA's Solar System
Workings program. Simulations has made use of: (i) the computing facilities
of the Laboratory of Astroinformatics (IAG/USP, NAT/Unicsul), supported
by FAPESP grant 2009/54006-4 and INCT-A, (ii) the Pleiades Supercomputer
of NASA's High-End Computing Capability, and (iii) the cluster of
the Department of Astronomy of the National Observatory of Rio de
Janeiro, acquired through CAPES grant 23038.007093/2012-13}

%%\bibliographystyle{apj}
%%\bibliography{jumping}

\newpage{}

\newpage{}

\begin{table}[H]
\protect\caption{Initial orbital elements of the five major planets. 
\label{tab:Planetary-initial-conditions}}

\centering{}\medskip{}
\begin{tabular}{ccccc}
\hline \hline 
Planet & Mass {[}$M_{\mathrm{Jup}}${]} & $a$ {[}au{]} & $e$ & $I$ {[}$^{\circ}${]}\tabularnewline
\hline 
Jupiter & 1.0 & 5.47 & 0.003 & 0.05\tabularnewline
Saturn & 0.299 & 7.45 & 0.011 & 0.02\tabularnewline
Ice \#1 & 0.053 & 10.11 & 0.017 & 0.11\tabularnewline
Ice \#2 & 0.053 & 16.08 & 0.006 & 0.07\tabularnewline
Ice \#3 & 0.053 & 22.17 & 0.002 & 0.05\tabularnewline
\hline \hline
\end{tabular}
\end{table}

\begin{table}[H]
\protect\caption{Statistic of surviving test particles during Phase 0.}
\label{statistic-of-surviving}

\centering{}\medskip{}
\begin{tabular}{cccc}
\hline \hline 
$t$ {[}Myr{]} & MB & HG & TS\tabularnewline
\hline 
10 & 6~820 & 992 & 44\tabularnewline
500 & 4~690 & 12 & 1\tabularnewline
\hline \hline 
\end{tabular}
\end{table}

\begin{table}[H]
\protect\caption{The median, mean and maximum dispersion of orbital elements (absolute
value of the difference between the elements at the beginning of Phase
1 and at the end of Phase 3), for each case of migration (EI set).}
\label{mean-dispersion}

\centering{}\medskip{}
\begin{tabular}{cccccccccc}
\hline \hline
 & \multicolumn{3}{c}{$\left|\delta a\right|$ {[}au{]}} & \multicolumn{3}{c}{$\left|\delta e\right|$} & \multicolumn{3}{c}{$\left|\delta I\right|$ {[}$^{\circ}${]}}\tabularnewline
 & Median & Mean & Max & Median & Mean & Max & Median & Mean & Max\tabularnewline
\hline 
Case 1 & 0.064 & 0.114 & 0.707 & 0.069 & 0.112 & 0.370 & 2.16 & 4.25 & 22.38\tabularnewline
Case 2 & 0.118 & 0.231 & 1.013 & 0.100 & 0.137 & 0.396 & 4.10 & 6.56 & 26.50\tabularnewline
Case 3 & 0.022 & 0.072 & 0.466 & 0.057 & 0.095 & 0.398 & 1.71 & 3.16 & 19.37\tabularnewline
\hline \hline
\end{tabular}
\end{table}

\end{document}